\documentclass[10pt]{iopart}

\usepackage[latin1]{inputenc}
\usepackage{iopams}
\usepackage{amsgen}
\usepackage{amsfonts}
\usepackage{amsbsy}
\usepackage{amssymb}
\usepackage{graphicx}
\usepackage{color}
\usepackage{enumitem}
\usepackage{xspace}
\usepackage[normalem]{ulem}
\usepackage{cancel}
\usepackage{hyperref}

\newcommand{\PRE}{Phys. Rev. E\xspace}
\renewcommand{\PRL}{Phys. Rev. Lett.\xspace}
\renewcommand{\EPL}{EPL\xspace}
\newcommand{\JSTATP}{J. Stat. Phys.\xspace}
\renewcommand{\JSTAT}{J. Stat. Mech.\xspace}
\renewcommand{\JMP}{J. Math. Phys.\xspace}
\renewcommand{\jpa}{J. Phys. A: Math. Theor.\xspace}

\newcommand{\titolo}{A framework for the direct evaluation of large deviations in non-Markovian processes}

\begin{document}

\title[Direct evaluation of large deviations in non-Markovian processes]{\titolo}

\author{Massimo Cavallaro$^{1,2}$ and Rosemary J.\ Harris$^{1}$}
\vspace{9pt}

\address{$^1$School of Mathematical Sciences, Queen Mary University of London, Mile End Road, London, E1 4NS, UK}
\vspace{2pt}
\address{$^2$School of Life Sciences and Department of Statistics, University of Warwick, Coventry, CV4 7AL, UK}

\vspace{9pt}
\eads{\mailto{m.cavallaro@warwick.ac.uk} and \mailto{rosemary.harris@qmul.ac.uk}}

\begin{abstract}
We propose a general framework to simulate stochastic trajectories with arbitrarily
long memory dependence and efficiently  evaluate large deviation functions associated to
time-extensive observables. This extends the ``cloning'' procedure of Giardin\'a \textit{et al.}~[Phys.~Rev.~Lett.~\textbf{96} 120603 (2006)] to non-Markovian systems.
We demonstrate the validity of this method by testing non-Markovian variants of an ion-channel model
and the Totally Asymmetric Exclusion Process, recovering results obtainable by other means.
 
\end{abstract}



\section{Introduction}
The theory of large deviations has been widely applied to equilibrium statistical mechanics~\cite{Ellis2006}.
Far from equilibrium, we can still deploy the same formalism, but targeting trajectories in space-time rather than static configurations~\cite{Merolle2005}. 
Instead of asking for the probability of observing a configuration with a given energy,
we require the probability of a trajectory of duration $T$ with a value $J$ of a time-extensive observable.
For this purpose, the details of the time evolution take on a major~role.

When the stochastic dynamics of a model system are Markovian, i.e., memoryless,
we can specify the rules for its evolution in time by means of the constant rates
$\beta_{x_j,x_i}$  of transitions from configuration $x_i$ to configuration $x_j$.
The full set of rates encodes inter-event times  with exponential waiting-time distributions,
which indeed possess the memoryless property.
However, to model real-world systems, such a simplified description may not be appropriate.
In fact, non-exponential waiting times seem to be relevant in many contexts, see, e.g.,~\cite{Ben-Avraham2005,Voit2005,Murray2007,Goh2008,Smith2011}.

Large deviation functionals have been computed in selected non-Markovian systems
(e.g., assuming the so-called~\textit{temporal additivity principle}~\cite{Harris2009},
or by defining hidden variables~\cite{Cavallaro2015}) and have revealed structure
hidden in the stationary state.
Analytical progress is difficult and simulations are necessary to explore
systematically more realistic models with memory.
However, to our knowledge, numerical schemes able to efficiently probe large deviation
functionals have been discussed in general
only for memoryless systems~\cite{Giardina2006,Lecomte2007,Gorissen2009,Nemoto2014,Nemoto2016}.
In this letter, we fill this literature gap and provide a general numerical method
to generate trajectories corresponding to arbitrarily rare values of $J$, based
on the ``cloning'' procedure of~\cite{Giardina2006}. The text is organised as follows.
In section~\ref{sec:Thermodynamics_of_trajectories}, we set up the general formalism, while in
section~\ref{sec:A_numerical_approach} we present the simulation scheme for
non-Markovian systems and the numerical method to evaluate large deviation functionals.
Section~\ref{sec:Semi-Markov_systems} deals with the special case of the semi-Markov process,
where the formalism has a particularly lucid interpretation in terms of the \textit{generalised Master equation}.
In section~\ref{sec:Examples} we test the method in some examples of increasing
complexity, where the large deviation functions can be computed exactly.
We conclude with a discussion in section~\ref{sec:discussion}.

\section{Thermodynamics of trajectories}
\label{sec:Thermodynamics_of_trajectories}
A \textit{trajectory} or \textit{history} of a stochastic process
starting at time $t_0$ and ending at time $t$,
on a configuration space $\mathcal{S}$, is defined as the 
sequence 
\begin{equation}
    w(t) := (t_0,x_0,~t_1,x_1,~t_2,x_2~\ldots~t_n,x_n,~t),
    \label{eq:trajectory}
\end{equation}
where $x_i \in \mathcal{S}$, $x_0$ is the initial configuration,
$t_0\le t_1 \le \ldots \le t$,
and $t_i$ (for $i=1,\ldots,n$) denotes the instant where the system jumps from configuration $x_{i-1}$ to $x_i$.
Specifically, we are interested in the probability density $\varrho [w(t)]$ that a trajectory $w(t)$ is observed.
Hereafter, we will consider a discrete configuration space $\mathcal{S}$,
although most of the arguments presented remain valid in the continuous case.
In general, we can separate $\varrho[w(t)]$ into a product over the contributions of each jump,
multiplied by the probability $P_{x_0}(t_0)$ that the configuration at $t_0$ is $x_0$, i.e.,
\begin{eqnarray}
    \varrho[w(t)] =& \phi_{x_n}[t-t_n;w(t_n)] \psi_{x_{n},x_{n-1}}[t_n-t_{n-1};w(t_{n-1})] \ldots \nonumber \\
     & \times \psi_{x_{1},x_{0}}[t_{1}-t_{0};w(t_0)] P_{x_0}(t_0) ,
     \label{eq:microcanonical}
\end{eqnarray}
where the generic factor  $\psi_{x_{n},x_{n-1}}[t_n-t_{n-1};w(t_{n-1})]$ is the
\textit{waiting-time probability density} (WTD)
that the transition from $x_{n-1}$ to $x_{n}$ occurs during the infinitesimal interval $[t_{n},t_{n}+\mathrm{d} t)$ given the history $w(t_{n-1})$;
it obeys the normalization $\sum_{x_n} \int_{t_{n-1}}^{\infty} \psi_{x_n ,x_{n-1}}[t_n-t_{n-1};w(t_{n-1})] \mathrm{d} t_n=1$.
Also, $\phi_{x_n}[t-t_n;w(t_n)] =  \sum_{x_n} \int_{t}^{\infty} \psi_{x_n, x_{n-1}}[t_n-t_{n-1};w(t_{n-1}) ] \mathrm{d} t_n$
is the survival probability of  the configuration $x_n$ for the interval $[t_n, t)$.
The special case without dependence on the history before the last jump will be considered in section~\ref{sec:Semi-Markov_systems}. 
The probability that the system has configuration $x \in \mathcal{S}$ at $t>t_0$ is
\begin{equation}
   P_{x}(t) = \sum_{n=0}^\infty \int_{t_0}^t \mathrm{d} t_1 \int_{t_1}^t \mathrm{d} t_2 \ldots \int_{t_{n-1}}^t \mathrm{d} t_n \sum_{x_0,x_1,\ldots,x_{n}} \delta_{x,x_n} \, \varrho  [w(t)] .
    \label{eq:Pxnt}
\end{equation}

We characterise a non-equilibrium system by means of time-extensive functionals $J[w(t)]$
of the trajectory $w(t)$, that can be written as the sum $\sum_{i=0}^{n-1} \theta_{x_{i+1},x_{i}}$
of elementary contributions corresponding to configuration changes.
Alternatively it is possible to consider ``static'' contributions
$\theta_{x_{i-1}}(t_i-t_{i-1})$. 
Functionals of these types may represent  physical observables integrated over the observation time $T=t-t_0$, e.g.,
the dynamical activity in a glassy system~\cite{Garrahan2009},
the moles of metabolites produced in a biochemical pathway~\cite{tomita2006metabolomics,Court2015},
the customers served  in a queuing network~\cite{stewart2009probability},
the flow in an interacting particle system~\cite{Derrida2007},
or certain quantities in stochastic thermodynamics~\cite{Harris2007,Seifert2012}.
Hereafter, we will refer to $J[w(t)]$ as the \textit{time-integrated current} and to its empirical
average $J[w(t)]/T=j(t)$ simply as \textit{current}.

The latter observable still fluctuates in time,
although it is doomed to converge to its ensemble average $\langle j \rangle$ in the limit $t \to \infty$ (with $t_0$ fixed and finite).
The consequent computational difficulty in obtaining the probability $\mathcal{P}(j,t)$ of having a given value of $j(t)$,
can be alleviated by introducing the ``canonical'' density $\e^{-s J[w(t)]} \varrho [w(t)]$, the ``partition function''
\begin{equation}
    Z(s,t) = \int  \e^{-s J[w(t)]} \varrho [w(t)]  \, \mathrm{d} w(t), \label{eq:partition_function}
\end{equation}
and the scaled cumulant generating function (SCGF)
\begin{equation}
    e(s) = -\lim_{t \to \infty}  \frac{1}{t} \ln Z(s,t),
    \label{eq:SCGF}
\end{equation}
where $s$ is an intensive conjugate field.
Assuming a large deviation principle, i.e.,
\begin{equation}
    \mathcal{P}(j,t) \sim \e^{-t \hat{e}(j)},
\end{equation}
the rate function $\hat{e}(j)$ can be obtained by a Legendre--Fenchel transform,
\begin{equation}
    \hat{e}(j) = \sup_s \{e(s) -s j\},
\end{equation}
when the SCGF is differentiable \cite{Touchette2009}.
Equation~\eref{eq:partition_function} can be written as
\begin{eqnarray}
    Z(s,t)  &= \sum_{n=0}^\infty \int_{t_0}^t \mathrm{d} t_1 \int_{t_1}^t \mathrm{d}t_2 \ldots \int_{t_{n-1}}^t \mathrm{d}t_n \sum_{x_0,x_1,\ldots,x_{n}} \nonumber \\
     & \phi_{x_n}[t-t_n;w(t_n)] \tilde{\psi}_{x_{n},x_{n-1}}[t_n-t_{n-1};w(t_{n-1})] \ldots \nonumber \\
     & \times    \tilde{\psi}_{x_{1},x_{0}}[t_{1}-t_{0};w(t_0)] P_{x_0}(t_0) , \label{eq:explicit_partition_function}
\end{eqnarray}
where
\begin{eqnarray}
\hspace*{-1cm}	\tilde{\psi}_{x_{n},x_{n-1}}[t_n-t_{n-1};w(t_{n-1})] 
	=	\e^{-s \theta_{x_{n},x_{n-1}}} \psi_{x_{n},x_{n-1}}[t_n-t_{n-1};w(t_{n-1})]
	\label{eq:biased_WTD}
\end{eqnarray}	
can be seen as the ``biased'' WTD of a stochastic dynamics that does not conserve total probability.
This immediately suggests that we can access the SCGF
by computing the exponential rate of divergence of an
ensemble of trajectories.

\section{A numerical approach}
\label{sec:A_numerical_approach}
\subsection{Non-Markovian stochastic simulation}
\begin{figure}
\begin{minipage}{0.2\textwidth}
{\color{white}.}
\end{minipage}
\begin{minipage}{0.8\textwidth}
\center
    \includegraphics[width=0.6\textwidth]{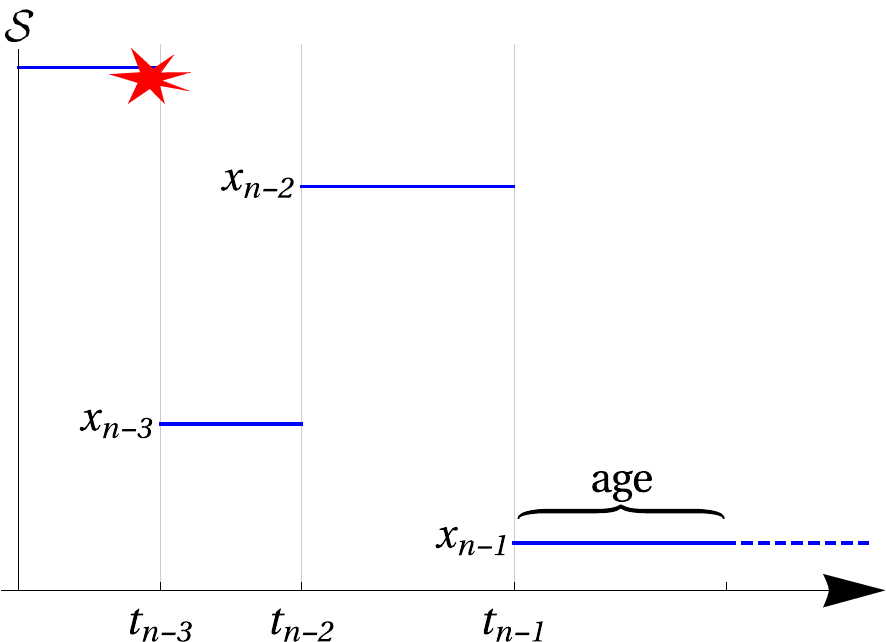}
\end{minipage}
    \caption{   \label{fig:figure1}
    Representation of a portion of a trajectory. The time elapsed from the last jump is called age.
    The conditional probability of having a configuration $x_n$ at an instant $t_n>t_{n-1}$ depends on the age,
    as well as on events which happened during the history (e.g., the  one marked by the red star).
    }
\end{figure}

The WTD can be expressed in terms of a time-dependent \textit{rate} or~\textit{hazard} $\beta_{x_n, x_{n-1}}[t_n-t_{n-1};w(t_{n-1})]$,
which is the probability density that there is a jump from $x_{n-1}$ to $x_{n}$ in $[t_n,t_n+ \mathrm{d} t)$,
conditioned on having no transitions during the interval~$[t_{n-1},t_{n})$, 
\begin{eqnarray}
    \psi_{x_n,x_{n-1}}[t_n-t_{n-1};w(t_{n-1})]  \nonumber \\
    =  \beta_{x_n, x_{n-1}}[t_n-t_{n-1};w(t_{n-1})] \times \phi_{x_{n-1}}[&t_n-t_{n-1};w(t_{n-1})].
    \label{eq:hazard}
\end{eqnarray}
For brevity we define $\tau =t_n-t_{n-1}$, which is the value of the \textit{age},
i.e. the time elapsed since the last jump,
at which the next jump takes place (see~\fref{fig:figure1}).
Roughly speaking, the hazard $\beta_{x_n, x_{n-1}}[\tau;w(t_{n-1})]$ is
the likelihood of having an almost immediate transition from a state $x_{n-1}$ known to be of age $\tau$,
to a state $x_n$, and, crucially, can also depend on the history $w(t_{n-1})$.
From equation~\eref{eq:hazard}, summing over $x_n \in \mathcal{S}$, and
defining $\beta_{x_{n-1}}[\tau;w(t_{n-1})] = \sum_{x_n} \beta_{x_{n},x_{n-1}}[\tau;w(t_{n-1})]$ and
$\psi_{x_{n-1}}[\tau;w(t_{n-1})] = \sum_{x_n} \psi_{x_n, x_{n-1}}[\tau;w(t_{n-1})]$, we get
\begin{eqnarray}
    \psi_{x_{n-1}}[\tau;w(t_{n-1})] & =-\frac{ \mathrm{d}}{\mathrm{d}\tau} \phi_{x_{n-1}}[\tau;w(t_{n-1})] \nonumber \\
&    =  \beta_{x_{n-1}}[\tau;w(t_{n-1})] \times \phi_{x_{n-1}}[\tau;w(t_{n-1})].
\label{eq:derivative_survival}
\label{eq:hazard_sum}
\end{eqnarray}
The age-dependent sum $\beta_{x_{n-1}}[\tau;w(t_{n-1})]$ is also referred to as
the \textit{escape rate} from $x_{n-1}$ and, using equation~\eref{eq:derivative_survival},
can be written as a logarithmic derivative
\begin{equation}
    \beta_{x_{n-1}} [\tau;w(t_{n-1})] = -\frac{\mathrm{d}}{ \mathrm{d} \tau} \ln  \phi_{x_{n-1}}[\tau;w(t_{n-1})].
    \label{eq:beta}
\end{equation}
Integrating equation~\eref{eq:beta} with initial condition $\phi_{x_{n-1}}[0;w(t_{n-1})] =1$ gives:
\begin{eqnarray}
\hspace*{-1cm}    \phi_{x_{n-1}}[\tau; w(t_{n-1}) ] &= \exp \left( - \int_{0}^{\tau} \beta_{x_{n-1}} [t;w(t_{n-1})] \mathrm{d} t \right)  ,  \label{eq:survival} \\
\hspace*{-1cm}    \psi_{x_{n-1}}[\tau; w(t_{n-1}) ] &= \beta_{x_{n-1}}[\tau;w(t_{n-1})] \exp \left( -\int_{0}^{\tau} \beta_{x_{n-1}} [t;w(t_{n-1})] \mathrm{d} t \right).
\end{eqnarray}
These let us cast equation~\eref{eq:hazard} in the more convenient form 
\begin{eqnarray}
     \psi_{ x_n, x_{n-1} } [\tau; w(t_{n-1})]  
     = p_{x_n,x_{n-1}}[\tau; w(t_{n-1}) ]  \times \psi_{x_{n-1}} [\tau;w(t_{n-1})],
     \label{eq:convenient_WTD}
\end{eqnarray}
where
\begin{equation}
	p_{x_n,x_{n-1}}[\tau; w(t_{n-1}) ] = \frac{\beta_{ x_n, x_{n-1} }[\tau;w(t_{n-1})]}{\sum_{x_{n}} \beta_{ x_n, x_{n-1}}[\tau ;w(t_{n-1})]  }
    \label{eq:probability_mass}
\end{equation}
is the probability that the system jumps into the state $x_n$,
given that it escapes the state $x_{n-1}$ at age $\tau$.
It is important to notice that the  normalization conditions $ \sum_{x_n} p_{x_n,x_{n-1}}[\tau; w(t_{n-1}) ] =1$ and
$\int_{0}^{\infty}  \psi_{x_{n-1}} [\tau;w(t_{n-1})] \mathrm{d} \tau=1$ are satisfied.
Hence, we can sample a random waiting time $\tau$, according to the density $\psi_{x_{n-1}} [\tau;w(t_{n-1})]$ and, after that,
a random arrival configuration $x_n$, according to the probability mass $p_{x_n,x_{n-1}}[\tau; w(t_{n-1}) ]$.
This suggests a standard Monte Carlo algorithm for the generation of a trajectory~\eref{eq:trajectory}:
\begin{enumerate}[label=\arabic*) ]
\item Initialise the system  to a configuration $x_0$ and a time $t_0$. Set a counter to~$n=1$.
\item  \label{item:item} Draw a value $\tau$ according to the density \eref{eq:hazard_sum} and update the time to~$t_n=t_{n-1}+\tau$.
\item Update the system configuration to $x_n$, with probability given by \eref{eq:probability_mass}.
\item Update $n$ to $n+1$ and repeat from~\ref{item:item} until $t_n$ reaches the desired simulation time.
\end{enumerate}

\subsection{The cloning step}
\label{sec:The_cloning_step}
We now need to take into account the effect of the factor $\e^{-s \theta_{x_{n},x_{n-1}}} $ on the dynamics,
which is to increment (if $\theta_{x_{n},x_{n-1}}<0$) or decrement (if $\theta_{x_{n},x_{n-1}}>0$) the ``weight'' 
of a trajectory, within an ensemble.
This can be implemented by means of the so-called ``cloning'' method,
which consists of assigning to each trajectory of the ensemble a population of identical clones proportional to its weight.
As reviewed, e.g., in~\cite{Grassberger2000,Grassberger2000coll}, the idea is not new.
It seems to be born in the context of quantum physics and can be traced back to Enrico Fermi~\cite{Metropolis1949}.
Noticeably, cloning was proposed as a general scheme for the evaluation of
large deviation functionals of non-equilibrium Markov processes in~\cite{Giardina2006}
(with further continuous-time implementation in~\cite{Lecomte2007}) and has also been
extensively applied within equilibrium statistical physics~\cite{Tailleur2007,JansevanRensburg2009,Hsu2011}.
Here, we refine and extend this idea for the case of non-Markovian processes.
One of the devices used in~\cite{Giardina2006,Lecomte2007} is to define modified
transition probabilities---valid \textit{only} under the Markovian assumption---and a
modified cloning factor, encoding the contraction or expansion of the trajectory weight.
In fact, it is implicit in the original work that the redefinition of such
quantities is unnecessary; in some cases it may also be inconvenient, see section~\ref{sec:Markovian_case}.
%
%
An arguably more natural choice, especially for non-Markovian dynamics, is to focus
on the WTDs. Specifically, equations \eref{eq:explicit_partition_function} and
\eref{eq:biased_WTD} suggest the following procedure:
\begin{enumerate}[label=\arabic*)]
\item Set up an ensemble of $N$ clones and initialise each with a given time $t_0$
, a random configuration $x_0$, and a counter $n=0$.
Set a variable $C$ to zero. For each clone, draw a time $\tau$ until the next jump from the density $\psi_{x_{0}}[\tau;w(t_0)]$,
and then choose  the clone with the smallest value of $t=t_0+\tau$.
\item For the chosen clone, update $n$ to $n+1$ and then the configuration from $x_{n-1}$ to $x_{n}$ according to the probability mass $p_{x_n,x_{n-1}}[\tau; w(t-\tau)]$.
\item Generate a new waiting time $\tau$ for the updated clone according to $\psi_{x_{n}}[\tau;w(t)]$ and increment its value of $t$ to $t+\tau$. 
\item \textbf{Cloning step.} \label{item:cloning_step} Compute $y = \lfloor \e^{-s\theta_{x_{n},x_{n-1}}} + u\rfloor$, where $u$ is drawn from a uniform distribution on $[0,1)$.
\begin{enumerate}[label=\arabic*)]
	\item If $y=0$, prune the current clone. Then replace it with another one, uniformly chosen among the remaining $N-1$.
	\item If $y>0$, produce  $y-1$  copies of the current clone.
    Then, prune a number $y-1$ of elements, uniformly chosen among the existing $N+y-1$.  
\end{enumerate}
\item Increment $C$ to $C + \ln [(N+\e^{-s \theta_{x_{n},x_{n-1}}}-1)/N]$.
Choose one of the clones with the smallest $t$, and repeat from~\ref{item:item} until $t-t_0$ for the chosen clone reaches the desired simulation time $T$.
\end{enumerate}
The SCGF is finally recovered as $-C/T$ for large $T$.
The net effect of step~\ref{item:cloning_step} is to maintain a constant population of samples
whose mean current does not decay to~$\langle j \rangle$.

\section{Semi-Markov systems}
\label{sec:Semi-Markov_systems}
A common framework for modelling systems with finite-range memory is the formalism of \textit{semi-Markov processes},
also referred to as \textit{continuous-time random walks} (CTRWs) in the physics literature.
CTRWs were introduced to model transport on lattices~\cite{Montroll1965,Haus1987} and later used in many other contexts,
e.g., to describe quantum dots~\cite{Esposito2008},
temporal networks~\cite{Hoffmann2012},
animal movements~\cite{Giuggioli2009,Al-Sabbagh2015},
biochemical reactions~\cite{Flomenbom2005},
and single-molecule kinetics~\cite{Qian2007,Wang2007}.
In such systems, the probability of having a jump depends only on the departure state and on the age,
meaning that the memory is lost after each jump and the dependence on the previous history is removed.
Under this assumption, the probability density of observing a trajectory~\eref{eq:trajectory} is
\begin{equation}
\hspace*{-0.5in}
   \varrho[w(t)] = \phi_{x_n}(t-t_n) \psi_{x_{n},x_{n-1}}(t_n-t_{n-1}) \ldots \psi'_{x_{1},x_{0}}(t_{1}-t_{0}) P_{x_0}(t_0),
    \label{eq:microcanonical_CTRW}
\end{equation}
where the primed WTD  can differ from the others%
\footnote{\label{fp} A natural situation is when we observe a portion of a trajectory that started before $t_0$.
In this case,  $\psi'_{x_{1},x_{0}}(t_{1}-t_{0}) = \psi_{x_{1},x_{0}}(t_{1}-t_{-1})/\phi_{x_{0}}(t_{0}-t_{-1})$,
as it depends on the time $t_{-1}$ of the last jump before $t_0$, being conditioned on the survival until $t_0$.
Also, in this case, 
the probability that $x_0$ survives the time
$t - t_0$ is $\phi'_{x_0}(t - t_0) = \phi_{x_0}(t - t_{-1})/\phi_{x_0}(t_0- t_{-1})$.
}.
When $p_{x_n,x_{n-1}}[\tau;w(t_{n-1})]$ has no dependence on $\tau$
(as well as, of course, no dependence on $w(t_{n-1})$, due to the semi-Markov dynamics),
we can write $\psi_{x_{n},x_{n-1}}(\tau) = p_{x_n,x_{n-1}}  \psi_{x_{n-1}} (\tau)$
and the process is said to satisfy direction-time independence (DTI)\footnote{As proved in~\cite{Andrieux2008}, for finite configuration space,
the DTI condition leads to the so-called Gallavotti--Cohen symmetry~\cite{Lebowitz1999},
which reduces to the time-reversal symmetry in the special case where the detailed balance condition is 
 also satisfied~\cite{Qian2007,Wang2007}.}.

In a slight shift in notation we now use $x_i$ and $x_j$ as configuration labels.
The probability $P_{x_i}(t)$ for the system to be in the configuration $x_i$ at time $t$,
in a semi-Markov process,
follows a convenient differential equation, see~\cite{supplemental} and references therein,
which is referred to as the generalised Master equation (GME):
\begin{equation}
\fl
    \frac{ \mathrm{d} }{\mathrm{d} t} P_{x_i}(t) = I_{x_i}(t-t_0) + \sum_{x_j \neq  x_i} \int_{t_0}^{t} \left[ K_{x_i,x_j}(t-\tau) P_{x_j}(\tau) - K_{x_j,x_i}(t-\tau)P_{x_i}(\tau) \right] \mathrm{d} \tau,
    \label{eq:generalised_Master}
\end{equation}
where $K_{x_i,x_j}(t-\tau)$ is the \textit{memory kernel},
taking into account the configuration at time $t-\tau$, and $I_{x_i}(t-t_0)$ depends on the primed WTDs~\cite{supplemental}.
The memory kernel is defined through an equation similar to $\eref{eq:hazard}$, but in the Laplace domain,
\begin{equation}
       \overline{\psi}_{x_i,x_j}(\nu) =  \overline{K}_{x_i,x_j}(\nu)  \overline{\phi}_{x_j}(\nu) ,
\end{equation}
with $\overline{f}(\nu) = \int_{0}^{\infty} \e^{-\nu T} f(T)  \mathrm{d} T $.

The statistics of time-extensive variables in semi-Markov processes have been studied in~\cite{Andrieux2008,Esposito2008,Maes2009}
and compared to memoryless processes.
In systems described by a standard Master equation, 
one strategy is to analyse  a process that obeys a modified rate equation,
obtained replacing  the time-independent rates $\beta_{x_i,x_j}$,
with the products $\e^{-s\theta_{x_i,x_j}}\beta_{x_i,x_j}$, which are referred to as ``biased'' rates.
This is particularly simple when $\theta_{x_i,x_j}$ can only take values $-1,0,1$, i.e.,
at each step, the total current varies,  at most, by one unit.
In semi-Markov systems it is possible to investigate the statistics of $J$ in a similar,
but more general, way. Instead of the standard Master equation, we deploy the GME~\eref{eq:generalised_Master}.
The probability $P_{(x_i,J)}(t)$ of having a configuration $x_i$ with total current $J$ at time $t$,
under the constraint that the current can only grow or decrease by one unit at each jump,
obeys the following GME:
\begingroup\makeatletter\def\f@size{9}\check@mathfonts
\begin{eqnarray}
\fl
    \frac{\mathrm{d}}{\mathrm{d}t} P_{(x_i,J)}(t) = I_{(x_i,J)}(t-t_0) + \sum_{x_j\neq x_i} \int_{t_0}^{t} K_{(x_i,J) \leftarrow (x_j,J)} (t-\tau) P_{(x_j,J)}(\tau) \mathrm{d} \tau \nonumber \\
\fl
    + \sum_{x_j} \int_{t_0}^{t} K_{(x_i,J) \leftarrow (x_j,J+1)} (t-\tau) P_{(x_j,J+1)}(\tau) \mathrm{d}\tau + \sum_{x_j} \int_{t_0}^{t} K_{(x_i,J) \leftarrow (x_j,J-1)}(t-\tau) P_{(x_j,J-1)}(\tau) \mathrm{d}\tau \nonumber \\
\fl
    - \sum_{x_j \neq x_i} \int_{t_0}^{t} K_{(x_j,J) \leftarrow (x_i,J)}(t-\tau)P_{(x_i,J)}(\tau)  \mathrm{d}\tau 
    - \sum_{x_j} \int_{t_0}^{t} K_{(x_j,J+1) \leftarrow (x_i,J)}(t-\tau)P_{(x_i,J)}(\tau) \mathrm{d}\tau \nonumber \\
\fl
    -  \sum_{x_j} \int_{t_0}^{t} K_{(x_j,J-1)\leftarrow (x_i,J)}(t-\tau)P_{(x_i,J)}(\tau) \mathrm{d}\tau
    \label{eq:generalised_Master_J_diag}
\end{eqnarray}
\endgroup
We now make the assumption that the memory kernels are independent of the 
total integrated current $J$ (only depending on the current increment),
i.e.,
\begin{equation}
        K_{(x_i,J) \leftarrow (x_j,J-c)}(t) = K_{x_i,x_j,c}(t),\\
    \label{eq:assumption}
\end{equation}
where $c=-1,0,1$.
The system is diagonalised with respect to the current subspace by means of the discrete Laplace
transform
\begin{equation}
	\tilde{P}_{x_i}(s,t) = \sum_J \e^{-sJ} P_{(x_i,J)}(t)	
\end{equation}
and is then equivalent to
\begin{eqnarray}
\fl    \frac{\mathrm{d}}{\mathrm{d}t}\tilde{P}_{x_i}(s,t) = \tilde{I}_{x_i}(s,t-t_0) + \sum_{x_j\neq x_i} \int_{t_0}^{t} K_{x_i,x_j,0}  (t-\tau) \tilde{P}_{x_j}(s,\tau) \mathrm{d} \tau      \nonumber \\
\fl
    + \sum_{x_j} \int_{t_0}^{t} \e^{s} K_{x_i,x_j,-1}(t-\tau) \tilde{P}_{x_j}(s,\tau) \mathrm{d} \tau +  \sum_{x_j}  \int_{t_0}^{t} \e^{-s} K_{x_i,x_j,+1}(t-\tau) \tilde{P}_{x_j}(s,\tau) \mathrm{d} \tau \nonumber \\
\fl
    - \sum_{c,x_j \neq x_i} \int_{t_0}^{t} K_{x_j,x_i,c}(t-\tau)\tilde{P}_{x_i}(s,\tau)  \mathrm{d} \tau \nonumber \\
\fl
    - \int_{t_0}^{t} K_{x_i,x_i,-1}(t-\tau) \tilde{P}_{x_i}(s,\tau)  \mathrm{d} \tau  - \int_{t_0}^{t}  K_{x_i,x_i,+1}(t-\tau) \tilde{P}_{x_i}(s,\tau)   \mathrm{d} \tau,
    \label{eq:generalised_Master_J_s_diag} 
\end{eqnarray}
which can be represented in a more compact form as 
\begin{equation}
	\partial_t | \tilde{P}(t)\rangle = \hat{\mathcal{L}}(t) | \tilde{P}(t) \rangle,
	\label{eq:compactest}
\end{equation}
where $\hat{\mathcal{L}}(t)$ is a linear $s$-dependent integral operator and
$|\tilde{P}(t)\rangle$ has components $\tilde{P}_{x_i}(s,t)$.
The limit as $t \to \infty$ of $\ln \langle 1| \tilde{P}(t) \rangle/t$ (where $\langle 1|$ is a row vector with all entries equal to one)
is the SCGF of $J$. Clearly, equation~\eref{eq:compactest}
does not conserve the product $\langle 1| \tilde{P}(t) \rangle$, except for $s=0$ when
this reduces to $\sum_{x_i} P_{x_i}(t)=1$.
The dynamics described by equation~\eref{eq:generalised_Master_J_s_diag}
is equivalent to the dynamics described by the GME~\eref{eq:generalised_Master_J_diag},
with the memory kernels, corresponding to jumps that contribute a unit $c$ in the 
total current, multiplied by a factor $\e^{-cs}$.
From linearity, it follows that the Laplace-transformed
kernels are
\begin{equation}
    \e^{-cs} \overline{K}_{x_i,x_j,c}(\nu)= \left. \e^{-cs} \overline{\psi}_{x_i,x_j,c}(\nu) \middle/ \overline{\phi}_{x_j}(\nu)\right.
    \label{eq:biased_laplace_kernel}
\end{equation}
This confirms that the modified dynamics can be simulated biasing the WTDs $\psi_{x_i,x_j,c}(t)$,
i.e., multiplying them by $\e^{-cs}$.

\label{sec:Markovian_case}
The Markovian case is recovered for $K_{x_i,x_j,c}(t) = \beta_{x_i,x_j,c}\delta(t)$.
Using this kernel, equations \eref{eq:generalised_Master_J_s_diag} and \eref{eq:compactest} can be written
as
\begin{equation}
    \partial_t | \tilde{P}(t) \rangle =  \tilde{G}  | \tilde{P}(t) \rangle,
\end{equation}
where $\tilde{G}$ is the $s$-modified stochastic generator of the Markov process
with time independent rates $\beta_{x_i,x_j}$ and components 
\begin{eqnarray}
\hspace{-1.7cm}
    \{ \tilde{G} \}_{x_i,x_j} = \beta_{x_i,x_j,0} +  \e^{-s} \beta_{x_i,x_j,+1} +  \e^{s} \beta_{x_i,x_j,-1},   \label{eq:KernelMatrix_} \\
\hspace{-1.7cm}
    \{ \tilde{G} \}_{x_i,x_i} =  \e^{-s} \beta_{x_i,x_i,+1}  +  \e^{s} \beta_{x_i,x_i,-1} - \beta_{x_i,x_i,-1}-\beta_{x_i,x_i,+1} - \beta_{x_i} ,
    \label{eq:KernelMatrix_diag}
\end{eqnarray}
where $\beta_{x_i} =  \sum_{c,x_j\neq x_i} \beta_{x_j,x_i,c}$ is the rate of escape from $x_i$%
\footnote{Note that the equation \eref{eq:KernelMatrix_diag}, and also the earlier~\eref{eq:generalised_Master_J_s_diag},
takes into account events that do not alter the configuration, but modify the current statistics.}.
%
This shows that  biasing the rates is consistent with biasing the WTDs (see also some related discussions in~\cite{Garrahan2009}).
However, from a numerical point of view, the latter choice remains convenient even for the Markovian case,
as it avoids us having to define the modified transition probabilities of~\cite{Lecomte2007}.
To see this, we consider the biased Markovian WTD
\begin{equation}
    \tilde{\psi}_{x_i,x_j,c}(\tau) = \e^{-cs} \beta_{x_i,x_j,c}  \exp \left(- \beta_{x_j}\tau  \right),
    \label{eq:biased_markov_WTD}
\end{equation}
which is the product of an exponential probability density
$\psi_{x_j}(\tau) = \beta_{x_j} \exp \left(-  \beta_{x_j} \tau \right)$,
a time-independent probability mass $p_{x_i,x_j,c} = \beta_{x_i,x_j,c} / \beta_{x_j}$,
and a simple cloning factor~$\e^{-cs}$.
These specify the two steps of the standard Doob-Gillespie algorithm for Markov processes~\cite{Gillespie1976},
followed by a cloning step of weight~$\e^{-cs}$.
Another legitimate choice is to
define $\tilde{\beta}_{x_i,x_j,c} = \e^{-cs}\beta_{x_i,x_j,c}$,
set $\tilde{\beta}_{x_j}= \sum_{x_i,c} \tilde{\beta}_{x_i,x_j,c} $, and write
\begin{equation}
 \tilde{\psi}_{x_i,x_j,c}(\tau) = \exp  \left[ \tau  \left( \tilde{\beta}_{x_j}  - \beta_{x_j} \right) \right] \tilde{\beta}_{x_i,x_j,c} \exp \left(- \tilde{\beta}_{x_j} \tau \right) .
\end{equation}
With such an arrangement, we recognise the algorithm of~\cite{Lecomte2007}, i.e., at each step,
the configuration evolves according to a stochastic generator with rates $\tilde{\beta}_{x_i,x_j,c}$,
and the ensemble is modified with the cloning factor $\exp  \left[ \tau  \left( \tilde{\beta}_{x_j}  - \beta_{x_j} \right) \right] $.
As the cloning factor here is exponential in time, during long intervals 
the relative number of new clones can be large.
This can cause major finite-ensemble errors, which are shown to be important, e.g.,  in~\cite{Hurtado2009,Cavallaro2015}.
Conversely,  an implementation  based on equation~\eref{eq:biased_markov_WTD}
seems to be one way to reduce (but not completely eliminate) such a problem.

\section{Examples}
\label{sec:Examples}
We now test our procedure against three non-Markovian models,
whose exact large deviations are known from  the literature
or can be deduced from Markovian models.

\subsection{Semi-Markov models for ion-channel gating with and without DTI}
\label{sec:Two_semi-Markov_models}
\begin{figure}
\begin{minipage}{0.5\linewidth}
\includegraphics[width=1\textwidth,trim={0 0 0 0cm},clip]{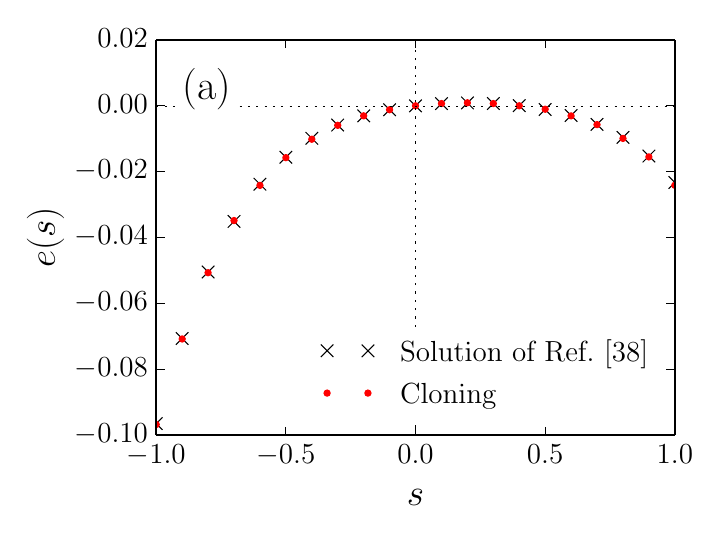}
\end{minipage}
\begin{minipage}{0.5\linewidth}
\includegraphics[width=1\textwidth]{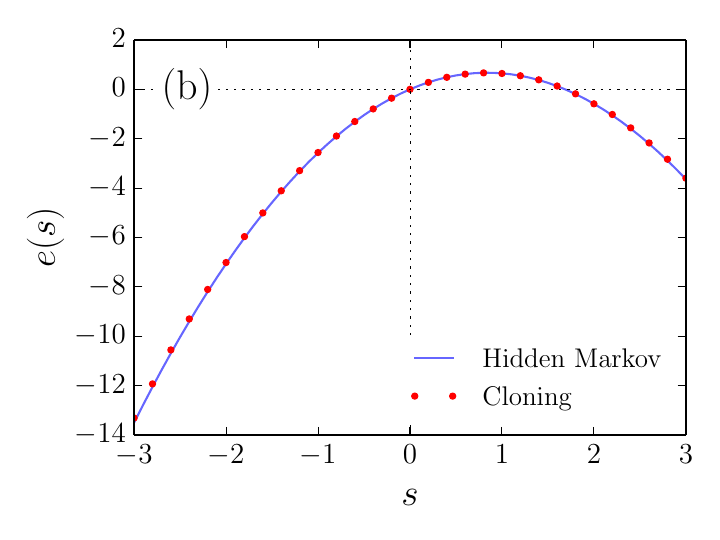}
\end{minipage}
\caption{\label{fig:two_state_model}
SCGF of current in ion channel. (a) DTI model with $(k_0,\lambda_0,k_1,\lambda_1)=(0.1,0.01,1,1)$
and $(p_{x_1,x_0,1},p_{x_0,x_1,0},p_{x_0,x_1,-1},p_{x_1,x_0,0})=(0.5,0.6,0.4,0.5)$;
the cloning result is consistent with the solution given in~\cite{Andrieux2008}.
(b) non-DTI model with Markov representation and inverse scales
$(\lambda^L_0,\lambda^R_0,\lambda^L_1,\lambda^R_1)=(10,10,20,20/3)$. 
The cloning reproduces the leading eigenvalue of the
Markovian $s$-modified generator.
In both cases $N=10^3$ and $T=10^3$.
}	
\end{figure}

The current through an ion channel in a cellular membrane can be modelled with
only two states, corresponding to the gate being singly occupied ($x_1$) or empty~($x_0$);
an ion can enter or leave this channel via the left ($L$) or right ($R$) boundary
and non-exponential waiting times lead to a complex behaviour~\cite{Barkai1996}. 
Specifically, we denote the WTD for a particle succeeding in entering
(or leaving) through the boundary
$L$ by $\psi_{x_1,x_0,1}(\tau)$ (or $\psi_{x_0,x_1,-1}(\tau)$)  with respective
density $\psi_{x_1,x_0,0}(\tau)$ (or $\psi_{x_0,x_1,0}(\tau)$) for the boundary $R$.
The rightwards current is measured by a counter that increases (decreases)
by one when a particle enters (leaves) the system through the boundary~$L$.
Its exact SCGF is obtained numerically in~\cite{Andrieux2008} as the leading pole of the time-Laplace
transform of $Z(s,t)$, for the DTI-case $\psi_{x_i,x_j,c}(\tau) = p_{x_i,x_j,c} \psi_{x_j}(\tau)$
with $\sum_{c=-1,0} p_{x_0,x_1,c}=\sum_{c=0,1} p_{x_1,x_0,c}=1$,
and the particular choice $\psi_{x_j}(\tau) = g(\tau;k_j,\lambda_j)$, where 
\begin{equation}
	g(\tau; k,\lambda)=\lambda^{k} \tau^{k-1} \exp ({-\lambda \tau})/\Gamma(k),
\end{equation}
with $\Gamma(k)$ as the Gamma function; the Markovian case is recovered for $k=1$.
Notably, the cloning method of~\sref{sec:A_numerical_approach}
can be implemented for any WTD, as only a bias of $\e^{s}$, for ions leaving the channel leftwards,
and a bias of $\e^{-s}$, for ions entering from left, are needed.
\Fref{fig:two_state_model}(a) shows that this method  reproduces, within numerical accuracy,
the solution given in~\cite{Andrieux2008}.

In a quest for more general classes of models to illustrate the power of our approach,
we now relax the constraint of DTI and assume that each 
transition can be triggered independently by two mechanisms, corresponding
to the two boundaries. We still assume that memory of the previous history is lost as soon
as the system changes state, thus preserving the semi-Markov nature.
At the instant when the gate is emptied, a particle attempts to enter the system from the left boundary
after a waiting time $T^L_{0}$ with density distribution $\bpsi^L_{0}(\tau)$, while another particle
attempts to arrive from the right boundary after a time $T^R_{0}$ distributed according to $\bpsi^R_{0}(\tau)$.
The waiting times $T^L_{1}$ and $T^R_{1}$, as well as the densities $\bpsi^L_{1}(\tau)$ and $\bpsi^R_{1}(\tau)$ are
defined similarly.
In order to have a right (left) jump during the interval $[\tau,\tau+\mathrm{d}\tau)$,
we also require that the left (right) mechanism remains silent until time $\tau$.
Consequently, the WTDs are
\begin{eqnarray}
	\psi_{x_1,x_0,0}(\tau) = \bpsi^{R}_{0}(\tau) \bvarphi^{L}_{0}(\tau), \qquad \psi_{x_1,x_0,1}(\tau)&= \bpsi^{L}_{0}(\tau) \bvarphi^{R}_{0}(\tau),\label{eq:semiMnoDITWTD1}\\
	\psi_{x_0,x_1,0}(\tau) = \bpsi^{R}_{1}(\tau) \bvarphi^{L}_{1}(\tau), \qquad \psi_{x_0,x_1,-1}(\tau) &= \bpsi^{L}_{1}(\tau) \bvarphi^{R}_{1} (\tau),\label{eq:semiMnoDITWTD2}
\end{eqnarray} 
where $\bvarphi^{(\rho)}_{j}(\tau) = \int_{\tau}^\infty \bpsi^{(\rho)}_{j}(t) \mathrm{d} t $
are survival probabilities, with $\rho$ denoting the mechanism $L$ or $R$.
As a concrete choice, we again assign a Gamma probability distribution to the waiting time of each event,
\begin{equation}
		\bpsi^{(\rho)}_{j}(\tau) = g(\tau; k^{(\rho)}_{j}, \lambda^{(\rho)}_{j}  ),
\end{equation}
so that the survival probabilities are
\begin{equation}
	\bvarphi^{(\rho)}_{j}(\tau)  =   {\Gamma(k^{(\rho)}_{j}, \lambda^{(\rho)}_{j} \tau)}/{\Gamma(k^{(\rho)}_{j})},
\end{equation}
where $\Gamma(k, x)$ is the upper incomplete Gamma function.
The time to the next jump, given that the system just reached state $x_j$ (i.e., its age is zero)
is $\min \{ T^L_{j} ,T^R_{j} \}$ and is  associated to the total survival probability,
\begin{equation}
	\phi_{x_j}(\tau) =  \bvarphi^{L}_{j}(\tau) \bvarphi^{R}_{j}(\tau).
	\label{eq:CDF0}
\end{equation}
Once the transition time is known, either the left or right trigger is chosen,
according to the age-dependent rates
\begin{equation}
	\beta^{(\rho)}_{j}(\tau) = { g(\tau; k^{(\rho)}_{j}, \lambda^{(\rho)}_{j} )\Gamma(k^{(\rho)}_{j})}/{ \Gamma(k^{(\rho)}_{j}, \lambda^{(\rho)}_{j} \tau)}.
\end{equation}
The SCGF of the left current is computed by biasing the WTDs $\psi_{x_1,x_0,1}(\tau)$
and $\psi_{x_0,x_1,-1}(\tau)$ with $\e^{\mp s}$, respectively.

While the implementation of the method of~\sref{sec:The_cloning_step} remains straightforward
for this model, a general solution for the exact SCGF is missing.
We thus specialise to the case with $k^R_0=k^R_1=k^L_0=k^L_1=2$,
for which the Laplace transform of $\psi_{x_i,x_j}(t)=\sum_c \psi_{x_i,x_j,c}(t)$ is a rational function of $\nu$, viz., 
\begin{eqnarray}
\hspace*{-1cm}    \overline{\psi}_{x_i,x_j}(\nu) = \left( \alpha_{j,2}^{R} + \alpha_{j,2}^{L}\right) \left( \frac{\lambda_j}{\nu+\lambda_j}\right)^2 +  \left( \alpha_{j,3}^{R} + \alpha_{j,3}^{L} \right) \left( \frac{\lambda_j}{\nu+\lambda_j}\right)^3,
\label{eq:semiMnoDTIWTD_Laplace}
\end{eqnarray}
where we defined for convenience $\lambda_j=\lambda^L_j+\lambda^R_j$ and
\begin{eqnarray}
\hspace*{-1cm}    \alpha^{L}_{j,2} = \frac{(\lambda^L_j)^2}{(\lambda_j)^2}, \quad  \alpha^{L}_{j,3} = \frac{2 (\lambda^L_j)^2 \lambda^R_j}{(\lambda_j)^3}  ,
\quad  \alpha^{R}_{j,2} = \frac{(\lambda^R_j)^2}{(\lambda_j)^2}, \quad  \alpha^{R}_{j,3} = \frac{2 (\lambda^R_j)^2 \lambda^L_j}{(\lambda_j)^3}.
\end{eqnarray}
This can be thought of as resulting from a walker spending exponentially distributed
times in three hidden stages, as in~\cite{cox1955}.
Hence, the two-state semi-Markov process with WTD~\eref{eq:semiMnoDITWTD1} and \eref{eq:semiMnoDITWTD2}
can be seen as a six-state Markov process, see~\cite{supplemental} for more details and an alternative formulation.
The linearity of the Laplace transform permits the distinction of the left and right
contributions in~\eref{eq:semiMnoDTIWTD_Laplace},
hence it remains easy to bias the rates that correspond to a change in $J$
and the SCGF can be exactly found as the leading eigenvalue of a
modified Markovian stochastic generator~\cite{supplemental}.
\Fref{fig:two_state_model}(b) shows convincing agreement of our method with this exact approach.

\subsection{Totally Asymmetric Exclusion Process (TASEP) with history dependence}
\label{sec:IPS_with_history-dependent_dynamics}
More general non-Markovian systems are those whose WTDs 
depend on events which occurred during the whole observation time.
Systems in this class are the ``elephant'' random walk~\cite{Schutz2004} and its analogues,
where the transition probabilities at time $t$ depend on the history through the time-averaged current $j(t)$.
We focus here on an interacting particle system  with such current-dependent rates, namely the TASEP of~\cite{Harris2015}.

Non-Markovian interacting particle systems can be described by assigning a trigger
for actual or attempted jumps 
with WTD $\bpsi_i[\tau;w(t)]$ and a corresponding survival function
$\bvarphi_i[\tau;w(t)]$ to each elementary event $i$ that controls the particle dynamics.
The probability density that the next event (or attempted event) is of type $i$ and occurs 
in the time interval $[t+\tau,t+\tau+\mathrm{d}t)$, given that, for each $j$, a time $\tau_j$ has elapsed since the last event of type $j$,
is given by%
\footnote{In a slight abuse of notation we continue to use the full $w(t)$ as a parameter
but explicitly show the conditioning on the $\tau_i$s.}
\begin{equation}
    \psi_i[\tau; w(t)] = \bpsi_i[\tau+\tau_i; w(t)|\tau_i ] \prod_{j \neq i} \bvarphi_j[\tau+\tau_j; w(t)|\tau_j],
\label{eq:wtd_IPS}
\end{equation}
where $\bpsi_i[\tau+\tau_i; w(t)|\tau_i ]=\bpsi_i[\tau+\tau_i; w(t)]/\bvarphi_i[\tau_i; w(t)]$ and $\bvarphi_i[\tau+\tau_i; w(t)|\tau_i]=\bvarphi_i[\tau+\tau_i; w(t)]/\bvarphi_i[\tau_i; w(t)]$.
With exact expressions for these WTDs, we can implement the algorithms of section~\ref{sec:A_numerical_approach}.

The TASEP consists of a one-dimensional lattice of length $L$, where each lattice
site $l$,  $1\le l \le L$, can be either empty ($\eta_l=0$) or occupied by a particle ($\eta_l=1$).
Particles on a site $l<L$ are driven rightwards: they attempt a bulk jump to site $l+1$ with WTD $\bpsi_{\mathrm{b}}[\tau;w(t)]$,
the attempt being successful if $\eta_{l+1}=0$, as in~\cite{Gorissen2012,Concannon2014,Khoromskaia2014}.
With open boundaries, a particle that reaches the rightmost site $L$  leaves the system with WTD $\bpsi_L[\tau;w(t)]$.
Also, as soon as $\eta_1 = 0$, a further boundary mechanism turns on for 
particles to arrive on the leftmost site after further time $\tau$ with WTD $\bpsi_0[\tau + \tau_0;w(t)|\tau_0]$.
The special choice $\bpsi_0[\tau;w(t)]=\alpha \e^{-\alpha \tau} $,
$\bpsi_{\mathrm{b}}[\tau;w(t)]= p \e^{-p \tau} $, and
$\bpsi_L[\tau;w(t)]=\beta \e^{-\beta \tau} $
corresponds to the standard Markovian TASEP with constant left, bulk, and right rates  $\alpha$, $p$, and $\beta$.

We now assume that only the left boundary has a non-exponential WTD,
while the particle triggers have exponential WTDs with rate $1$ for
free particles in the bulk, and rate $\beta$ for the particle on the rightmost site.
Consequently the inter-event time density distribution, conditioned on a time $\tau_0$
having elapsed since the last attempted arrival, is 
\begin{eqnarray}
\hspace*{-1cm}    \psi[\tau;w(t)|\tau_0] &= \left( \frac{\bpsi_0 [\tau+\tau_0; w(t)]}{\bvarphi_0[\tau+\tau_0;w(t)]}(1-\eta_1) + \mathsf{n} +\beta \eta_L \right) \nonumber \\
\hspace*{-1cm} &\times \exp \left\{ \ln \left[ \frac{\bvarphi_0[\tau+\tau_0;w(t)]}{\bvarphi_0 [\tau_0;w(t)]} \right] (1-\eta_1) -(\mathsf{n} + \beta \eta_L) \tau \right\} .
    \label{eq:w_i_ASEP_0}
\end{eqnarray}
The probability mass distribution, conditioned on an age $\tau$ and elapsed time $\tau_0$ is
\begin{eqnarray}
    p_0[\tau;w(t)|\tau_0] &=  \frac{\bpsi_0[\tau+\tau_0;w(t)]}{\bvarphi_0[\tau+\tau_0;w(t)]}(1-\eta_1) \nonumber \\
    &\times \left( \frac{\bpsi_0[\tau+\tau_0;w(t)]}{\bvarphi_0[\tau+\tau_0;w(t)]}(1-\eta_1) + \mathsf{n} + \beta \eta_L \right)^{-1} ,\\
    p_i[\tau;w(t)|\tau_0] &=  \left( \frac{\bpsi_0[\tau+\tau_0;w(t)]}{\bvarphi_0[\tau+\tau_0;w(t)]}(1-\eta_1) + \mathsf{n} + \beta \eta_L \right)^{-1},\\
    p_L[\tau;w(t)|\tau_0] &=  \beta \eta_L \left( \frac{\bpsi_0[\tau+\tau_0;w(t)]}{\bvarphi_0[\tau+\tau_0;w(t)]}(1-\eta_1) + \mathsf{n} + \beta \eta_L \right)^{-1},
\label{eq:w_i_ASEP_3}
\end{eqnarray}
where $\eta_1$ and $\eta_L$ encode the exclusion rules of the TASEP, $i=1,2,\ldots,\mathsf{n}$,
and $\mathsf{n}$ is the number of free particles in the bulk, which depends on the lattice configuration before the jump.

\begin{figure}
    \includegraphics[width=1\textwidth]{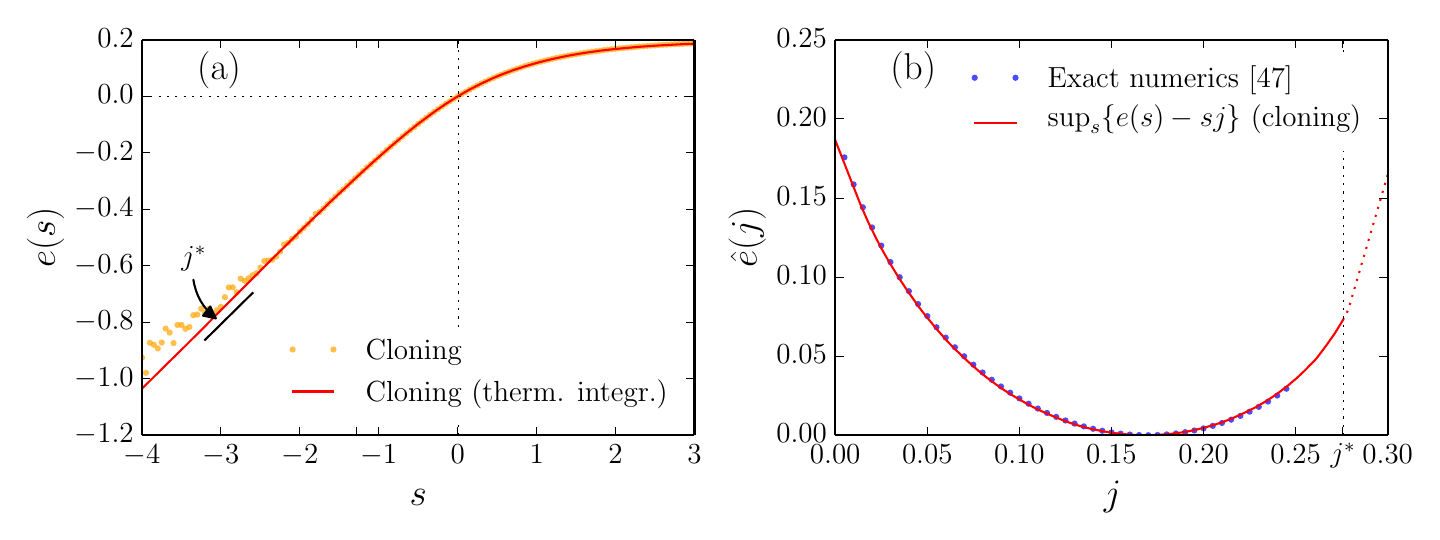}
    \caption{   \label{fig:mem_ASEP}
    (a) Cloning evaluation of the SCGF for the non-Markovian TASEP, with $(\alpha_0,a,\beta,L)=(0.2,0.1,1,10^3)$, using $T=10^3$.
    Ensemble size is $N=5\cdot 10^3$ ($N=10^4$) for $s>-2$ ($s<-2$).
    The markers correspond  to the direct evaluation of $e(s)$.
    Numerical errors are of the order of the symbol size, except for large negative $s$, 
    where finite-ensemble effects still seem to play a role, as documented in~\cite{Hurtado2009}.
    The red line is obtained as $\int_0^s ( \mathrm{d} e(\sigma)/\mathrm{d}\sigma) \, \mathrm{d} \sigma $,
    according to the thermodynamic integration of~\cite{Lecomte2007}.
    (b) Comparison between the Legendre--Fenchel transform of the red line in (a) and the rate function of~\cite{Harris2015}.
    The dotted line is a numerical artefact due to the finite range of $s$ in (a); the Legendre--Fenchel transform maps the whole linear branch of $e(s)$ to
    the value at $j^*$ and larger values of $j$ are, in fact, not probed.}
\end{figure}

Let us impose now that the arrival rate $\alpha$ depends linearly
on the input current $j(t)$, i.e., $\alpha(j) = \alpha_0 + a  j $,
which defines a time-dependent rate $\beta_0(t) := \alpha[j(t)]$.
Similar functional dependence (but on the instantaneous output current) has been used to model ribosome recycling in protein translation~\cite{Gilchrist2006,Sharma2011}.
Generically, such rates describe a simple form of positive feedback (for $a> 0$), whose effect on the stationary state of the TASEP is to shrink the \textit{low-density} phase~\cite{Harris2015,Sharma2011}.
The current fluctuations are also altered;
the rate function $\hat{e}(j)$ in this phase has already been computed, for our model, by means
of the temporal additivity principle~\cite{Harris2009,Harris2015}, hence this model provides a testing ground for the cloning method of~\Sref{sec:A_numerical_approach}.
The particle arrival mechanism activates 
when the leftmost site is emptied, when we set an age of $\tau=0$.
Denoting by $q$ the current immediately after the last arrival, which occurred at $t-\tau'$,
the value $j(t+\tau)$ at age $\tau$ can be expressed as $q \, (t-\tau')/(t+\tau)$,
hence the trigger hazard is
\begin{equation}
    \beta_0[\tau+\tau_0;w(t)] = 
    \alpha_0 + a \, q \, (t-\tau')/(t+\tau).
\end{equation}
Initial values of $\tau'$ and $q$ are chosen to 
be $1$ and $0$, respectively. This allows us to derive the trigger survival probability
and trigger WTD (see also~\cite{Harris2015}) which are, respectively,
\begin{eqnarray}
    \bvarphi_0[\tau_0+\tau;w(t)|\tau_0] &= \left( \frac{t}{t+\tau} \right)^{a \, q \, (t-\tau')} \e^{-\alpha_0 \tau}, \label{eq:survpmem}
\end{eqnarray}
and $\bpsi_0[\tau+\tau_0;w(t)] = \beta_0[\tau+\tau_0;w(t)]\bvarphi_0[\tau+\tau_0;w(t)]$.
Using these in equations~\eref{eq:w_i_ASEP_0} and \eref{eq:w_i_ASEP_3} allows us
to generate the trajectories, and the large deviation function can be evaluated by
applying a bias $\e^{-s}$ to the arrival WTD
and following the algorithm of~\sref{sec:A_numerical_approach}. The results are plotted in~\fref{fig:mem_ASEP}
and validated by the exact numerical calculation of~\cite{Harris2015}  (which assumes the temporal additivity principle).

It is worth noting that, for large negative values of $s$,
the SCGF displays a linear branch with slope $j^*$. 
If $e(s)$ remains linear with the same slope for $s<-4$, its Legendre--Fenchel transform will be defined only for $j \le j^*$.
This appears to be related to the dynamical phase transition seen in the Markovian TASEP, where large current fluctuations 
require correlations on the scale of the system size and the rate function, in the corresponding regime,
diverges with $L$~\cite{Lazarescu2015}.
Indeed, space-time diagrams (not shown) of the density profile
from the cloning simulations seem to suggest that the correlation
length increases as $s$ becomes more negative.
It would be interesting to further reduce the finite-ensemble errors
(as discussed in section~\ref{sec:discussion})
in order to probe larger negative values of~$s$.

\section{Discussion}
\label{sec:discussion}

We have demonstrated that the ``cloning'' algorithm for
the evaluation of large deviations can be applied consistently
for both Markovian \textit{and} non-Markovian dynamics.
In fact, the cloning/pruning of trajectories at each temporal step can be performed
according to a very simple factor multiplying the WTDs, as in equation~\eref{eq:biased_WTD}.
Our analysis encompasses classes of systems with different memory dependence and exploits
the similarities between their different formalisms.
The efficacy of this approach is confirmed by numerical results for some
of the rare non-Markovian models whose large deviation
functions can be obtained exactly.

For general non-Markovian cases, the implementation of our procedure is
not much harder than the exact simulation of the original trajectories.
In Markov processes, the procedure is equivalent
to those of \cite{Giardina2006,Lecomte2007},
 where biased dynamics involving alternative rates or transition probabilities have been defined. 
We expect that, to minimize finite-ensemble effects, an optimal choice of modified
WTDs and cloning factors exists for both non-Markovian and Markovian systems,
along the lines of the feedback control of~\cite{Nemoto2016}.
Further developments can thus be anticipated.

We also mention that the discrete-time case of~\cite{Giardina2006} is
interesting as the jumps and the cloning steps occur simultaneously
for each ensemble element.
This feature can be used to prevent a single clone
replacing a macroscopic fraction of the ensemble,
thus reducing finite size effects~\cite{supplemental}.
In continuous time an equivalent strategy is to mimic the discrete-time steps, as in,~e.g.,~\cite{Berryman2010},
so each trajectory evolves independently for a constant interval $\Delta t$;
in this case, the product of the cloning factors encountered during the interval, as well as the time elapsed since the last jump must be stored. This permits the application of the cloning step to all clones simultaneously.

Large deviation functionals are often hard to obtain analytically,
and such a difficulty is exacerbated in non-Markovian systems,
which better describe real-world situations;
we believe that the results of this work
open up a promising avenue for numerical studies.

\ack
It is a pleasure to thank Ra\'ul J. Mondrag\'on, Stefan Grosskinsky,
Oscar Bandtlow, and Arturo Narros for many helpful discussions.

\section*{Software availability}
The software supporting this study is publicly available at \href{https://github.com/mcavallaro/cloning}{github.com/mcavallaro/cloning}.

\section*{References}
\bibliographystyle{unsrt}

\pagebreak
\begin{center}
\textbf{\large Supplementary data for \\``\titolo ''}
\end{center}
\author{Massimo Cavallaro$^{1,2}$ and Rosemary J.\ Harris$^1$}
\vspace{9pt}
\address{$^1$School of Mathematical Sciences, Queen Mary University of London, Mile End Road, London, E1 4NS, UK}
\vspace{2pt}
\address{$^2$School of Life Sciences and Department of Statistics, University of Warwick, Coventry, CV4 7AL, UK}
\vspace{9pt}
\eads{
\mailto{m.cavallaro@warwick.ac.uk} and 
\mailto{rosemary.harris@qmul.ac.uk}
}
\setcounter{equation}{0}
\setcounter{figure}{0}
\setcounter{table}{0}
\setcounter{page}{1}
\setcounter{section}{0}
\makeatletter
\renewcommand{\theequation}{S\arabic{equation}}
\renewcommand{\thefigure}{S\arabic{figure}}

\section{Generalised master equation and its long-time behaviour}

\label{sec:Derivation_of_the_generalised_Master_equation}
We first derive the generalised master equation of a continuous-time random walk along the lines of~\cite{SMontroll1965,SHaus1987,SEsposito2008}.
For a semi-Markov process (represented by equation~\eref{eq:microcanonical_CTRW} in the main text)
the probability of having a configuration $x$ at time $t$, analogue of the equation~\eref{eq:Pxnt} in the main text,
can be explicitly written as the sum
\begin{eqnarray}
\fl   P_{x}(t) = \sum_{x_0} \delta_{x,x_0} \phi'_{x_0}(t - t_0) P_{x_0}(t_0)
+ \int_{t_0}^{t} \mathrm{d} t_1 \sum_{x_0, x_1} \delta_{x,x_1} \phi_{x_1}(t - t_1) \psi'_{x_1,x_0}(t_1 - t_0)  P_{x_0}(t_0)     \nonumber \\
\fl  + \int_{t_0}^{t} \mathrm{d} t_1 \int_{t_1}^{t} \mathrm{d}t_2 \sum_{x_0, x_1, x_2} \delta_{x,x_2} \phi_{x_2}(t - t_2)  \psi_{x_2,x_1}(t_2 - t_1) \psi'_{x_1,x_0}(t_1 - t_0) P_{x_0}(t_0)    \nonumber \\
\fl  + \int_{t_0}^{t} \mathrm{d} t_1 \int_{t_1}^{t} \mathrm{d} t_2 \int_{t_2}^{t}  \mathrm{d} t_3 \sum_{x_0, x_1, x_2, x_3} \delta_{x,x_3} \nonumber \\
\fl \quad \times \phi_{x_3}(t-t_3)\psi_{x_3,x_2}(t_3 - t_2)  \psi_{x_2,x_1}(t_2 - t_1) \psi'_{x_1,x_0}(t_1 - t_0) P_{x_0} (t_0) \,  +   \ldots \, ,
\label{eq:explicitPxnt}
\end{eqnarray}
where ``moves'' to the same configuration as the departure one are obviously excluded, i.e.,
$x_1\neq x_{0},x_2\neq x_{1}, \ldots$.
Denoting by $\eta_{x_i}(T)$ the probability that the system jumps onto the state
$x_i$ after a time $T=t-t_0$ since a reference instant $t_0$
(such a probability is also the sum of all the arguments of the integral operator
$\int_{t_0}^{t} \mathrm{d} \tau \phi_{x}(t - \tau) \cdot$ in equation~\eref{eq:explicitPxnt}),
we can write the standard recursive relations~\cite{SMontroll1965,SEsposito2008}
\begin{eqnarray}
\hspace*{-1.5cm}
P_{x_i}(T+t_0) &=   \phi'_{x_i}(T) P_{x_i}(t_0)  +   \int_{0}^{T} \phi_{x_i} (T-u) \eta_{x_i}(u) \, \mathrm{d} u,  \label{eq:std1}\\
\hspace*{-0.75cm}
\eta_{x_i}(T) &= \sum_{x_j \neq x_i} \psi'_{x_i,x_j}(T) P_{x_j}(t_0) +  \sum_{x_j\neq x_i} \int_{0}^{T} \psi_{x_i,x_j} (T-u) \eta_{x_j}(u) \, \mathrm{d} u,   \label{eq:std2}
\end{eqnarray}
where $x_i$ and $x_j$ are now generic configuration labels.
Equations \eref{eq:std1} and \eref{eq:std2} can be expressed even more compactly after a Laplace transform, i.e.,
\begin{eqnarray}
    \widehat{P}_{x_i}(\nu) &=  \overline{\phi'}_{x_i}(\nu) P_{x_i}(t_0)  +  \overline{\phi}_{x_i} (\nu) \overline{\eta}_{x_i}(\nu) , \label{Seq:rough_bar_P} \\
    \overline{\eta}_{x_i}(\nu) &= \sum_{x_j\neq x_i} \overline{\psi'}_{x_i,x_j}(\nu) P_{x_j}(t_0) + \sum_{x_j\neq x_i} \overline{\psi}_{x_i,x_j}(\nu) \overline{\eta}_{x_j}(\nu), \label{Seq:rough_bar_eta}
\end{eqnarray}
where $\overline{f}(\nu) = \int_{0}^{\infty} \e^{-\nu T} f(T)\, \mathrm{d} T$,  $\nu$ is the variable conjugated to $T$ and $\widehat{f}(\nu) = \int_{0}^{\infty} \e^{-\nu T} f(T+t_0)\, \mathrm{d} T$.
Using the explicit form  for the Laplace transform of the survival probabilities~\cite{Scox1962renewal},
\begin{equation}
    \overline{\phi}_{x_i} (\nu)  = \frac{1-\overline{\psi}_{x_i}(\nu)}{\nu}, \qquad \overline{\phi'}_{x_i} (\nu)  = \frac{1-\overline{\psi'}_{x_i}(\nu)}{\nu},
    \label{eq:survival_laplace}
\end{equation}
we get, from equation~\eref{Seq:rough_bar_P},
\begin{equation}
\hspace*{-1.7cm}
\nu \widehat{P}_{x_i} (\nu) - P_{x_i}(t_0)= -\sum_{x_j\neq x_i} \overline{\psi'}_{x_j,x_i}(\nu)  P_{x_i}(t_0) +  \overline{\eta}_{x_i}(\nu)  - \sum_{x_j \neq x_i} \overline{\psi}_{x_j,x_i}(\nu)  \overline{\eta}_{x_i}(\nu).
\end{equation}
Then, using~\eref{Seq:rough_bar_eta} to substitute for the second term of the r.h.s., yields
\begin{eqnarray}
\hspace*{-1.7cm}
\nu \widehat{P}_{x_i} (\nu) -  P_{x_i}(t_0) =& -  \sum_{x_j \neq x_i } \overline{\psi'}_{x_j,x_i}(\nu) P_{x_i}(t_0)  + \sum_{x_j \neq x_i} \overline{\psi'}_{x_i,x_j}(\nu) P_{x_j}(t_0)\nonumber \\
&+ \sum_{x_j \neq x_i} \overline{\psi}_{x_i,x_j}(\nu) \overline{\eta}_{x_j}(\nu) - \sum_{x_j\neq x_i} \overline{\psi}_{x_j,x_i}(\nu) \overline{\eta}_{x_i}(\nu) .\label{Seq:ci_siamo_quasi}
\end{eqnarray}
Plugging $\overline{\eta}_{x_i}(\nu)$ from equation~\eref{Seq:rough_bar_P} into
the third and fourth terms on the r.h.s.\ of equation~\eref{Seq:ci_siamo_quasi},
we get the equation
\begin{equation}
\hspace*{-1.7cm}
\nu \widehat{P}_{x_i} (\nu) -  P_{x_i}(t_0) = \overline{I}_{x_i}(\nu) +  \sum_{x_j\neq x_i}  \frac{\overline{\psi}_{x_i,x_j}(\nu)}{\overline{\phi}_{x_j}(\nu)}  \widehat{P}_{x_j}(\nu) -  \sum_{x_j\neq x_i} \frac{\overline{\psi}_{x_j,x_i}(\nu) }{\overline{\phi}_{x_i}(\nu)}  \widehat{P}_{x_i}(\nu),
 \label{Seq:ci_siamo}
\end{equation}
where $\overline{I}_{x_i}(\nu)$ contains the terms that explicitly depend on the initial conditions, i.e.,
\begin{eqnarray}
\hspace*{-1.6cm}  \overline{I}_{x_i}(\nu)=
 & - \sum_{x_j\neq x_i} \overline{\psi'}_{x_j,x_i}(\nu) P_{x_i}(t_0)
   + \sum_{x_j\neq x_i} \overline{\psi'}_{x_i,x_j}(\nu) P_{x_j}(t_0)\nonumber \\
\hspace*{-1.6cm} & - \sum_{x_j\neq x_i} \overline{\psi}_{x_i,x_j}(\nu) \frac{\overline{\phi'}_{x_j}(\nu)}{\overline{\phi}_{x_j}(\nu)}P_{x_j}(t_0)
   + \sum_{x_j\neq x_i} \overline{\psi}_{x_j,x_i}(\nu) \frac{\overline{\phi'}_{x_i}(\nu)}{\overline{\phi}_{x_i}(\nu)}  P_{x_i}(t_0).
\end{eqnarray}
After an inverse Laplace transform of equation~\eref{Seq:ci_siamo},
using the formula for the Laplace transform of a derivative on the l.h.s.,
we readily get the generalised master equation~\eref{eq:generalised_Master} of the main text.

In some situations, it is not necessary to deal with the initial-condition term.
As an obvious example, if the trajectory begins at the instant where a transition occurs
(as in the cases of section~\ref{sec:Two_semi-Markov_models}),
then $\overline{\psi}_{x_i,x_j}(\nu)=\overline{\psi'}_{x_i,x_j}(\nu)$
and $\overline{I}_{x_i}(\nu)=0$.
The same cancellation occurs when we observe a portion of a trajectory
that started before $t_0$ with exponential WTDs
(see footnote on page \pageref{fp} of the main text).
Focusing on the long-lime behaviour, 
it is also possible to prove that, for many other natural choices of the WTDs, 
\begin{equation}
  \lim_{t\to \infty} I_{x_i} (t) = 0. 
\end{equation}
To see this, we follow~\cite{SShlesinger1978,SEsposito2008} and consider WTDs that
have only finite moments, so that the following Maclaurin series expansion converges:
\begin{eqnarray}
\hspace*{-1.6cm} \overline{\psi}_{x_j,x_i}(\nu) &= \int_0^\infty \e^{-\nu \tau} {\psi}_{x_j,x_i}(\tau) \, \mathrm{d} \tau \nonumber \\
\hspace*{-1.6cm} &= \int_0^\infty {\psi}_{x_j,x_i}(\tau) \mathrm{d}\tau - \nu \int_0^\infty \tau {\psi}_{x_j,x_i}(\tau) \mathrm{d}\tau+\frac{\nu^2}{2}\int_0^\infty \tau^2 {\psi}_{x_j,x_i}(\tau) \mathrm{d}\tau+\ldots \nonumber \\
\hspace*{-1.6cm} &= P_{x_j,x_i} - \nu A_{x_j,x_i} + O(\nu^2) ,
\label{eq:WTD_alpha_1}
\end{eqnarray}
where the $P_{x_j,x_i}$ and $A_{x_j,x_i}$ are, respectively,
the zeroth and first moments of ${\psi}_{x_j,x_i}(\tau)$, in this case.
Alternatively, we consider $\alpha$-stable distributions, defined by their Laplace transform
\begin{eqnarray}
    \overline{\psi}_{x_j,x_i}(\nu) &=  P_{x_j,x_i} \exp(-\nu^\alpha B_{x_j,x_i}/P_{x_j,x_i}) \nonumber \\
      &= P_{x_j,x_i} - \nu^\alpha B_{x_j,x_i} + O( \nu^{2\alpha} )  ,
      \label{eq:WTD_alpha_stable}
\end{eqnarray}
where $P_{x_j,x_i}$ and $B_{x_j,x_i}$ are implicitly defined after expanding $\exp(-\nu^\alpha B_{x_j,x_i}/P_{x_j,x_i})$.
Here $0<\alpha < 1$ which corresponds to WTDs that, in the time domain, decay
as $\sim t^{-\alpha-1}$ and have infinite mean waiting times.
In both cases \eref{eq:WTD_alpha_1} and \eref{eq:WTD_alpha_stable}, the limits as $\nu \to 0$ of 
$\overline{\psi}_{x_i,x_j}(\nu)$ and $\overline{\psi'}_{x_i,x_j}(\nu)$
can be represented by the algebraic forms
$P_{x_i,x_j} - B_{x_i,x_j} \nu^\alpha $ and $P'_{x_i,x_j} - B'_{x_i,x_j} \nu^\alpha $, respectively.
Using the standard relations~\eref{eq:survival_laplace} and setting $B_{x_j} = \sum_{x_i\neq x_j} B_{x_i,x_j}$ and 
$B'_{x_j} = \sum_{x_i\neq x_j} B'_{x_i,x_j}$,
we get
\begin{eqnarray}
\fl
\lim_{\nu \to 0} \overline{I}_{x_i}(\nu) = 
\lim_{\nu \to 0} \sum_{x_j\neq x_i} \bigg[ -\left( P'_{x_j,x_i} -B'_{x_j,x_i} \nu^\alpha \right) P_{x_i}(t_0)
   +  \left( P'_{x_i,x_j} -B'_{x_i,x_j} \nu^\alpha \right) P_{x_j}(t_0)   \nonumber \\
\hspace*{-1cm}
    -  \left( P_{x_i,x_j} -B_{x_i,x_j} \nu^\alpha\right) \frac{B'_{x_j}}{B_{x_j}}  P_{x_j}(t_0)
   +  \left( P_{x_j,x_i} -B_{x_j,x_i} \nu^\alpha \right) \frac{B'_{x_i}}{B_{x_i}} P_{x_i}(t_0)\bigg] ,
\end{eqnarray}
which is  finite and implies, by the final value theorem, 
$\lim_{t\to \infty} I_{x_i} (t) = \lim_{\nu \to 0} \nu \overline{I}_{x_i}(\nu) =0$.
This suggests that, often, we do not need to know the exact behaviour of $I_{x_i} (t)$
to investigate the long-time limit of the generalised master equation.

We now consider the $s$-dependent case, and study the asymptotic behavior of
$\tilde{I}_{x_i} (s,t) = \sum_J \e^{-s J} I_{(x_i,J)}(t) $. In the joint configuration-current
space, the term encoding for the initial WTDs is
\begin{eqnarray}
\fl  \overline{I}_{(x_i,J)}(\nu)=
\sum_{x_j \neq x_i} \overline{\psi'}_{x_i,x_j,0}(\nu) P_{(x_j,J)}(t_0) 
   + \sum_{x_j,c=\pm 1}\overline{\psi'}_{x_i,x_j,c}(\nu) P_{(x_j,J-c)}(t_0)\nonumber \\
\fl\quad -\left[   \sum_{x_j \neq x_i}\overline{\psi'}_{x_j,x_i,0}(\nu)
   + \sum_{x_j,c=\pm 1}\overline{\psi'}_{x_j,x_i,c}(\nu) \right]  P_{(x_i,J)}(t_0) \nonumber \\
\fl \quad  + \sum_{x_j \neq x_i}\overline{\psi}_{x_j,x_i,0}(\nu) \frac{\overline{\phi'}_{x_i}(\nu)}{\overline{\phi}_{x_i}(\nu)} P_{(x_i,J)}(t_0)
   + \sum_{x_j,c=\pm 1}\overline{\psi}_{x_j,x_i,c}(\nu) \frac{\overline{\phi'}_{x_i}(\nu)}{\overline{\phi}_{x_i}(\nu)} P_{(x_i,J-c)}(t_0) \nonumber \\
\fl \quad -\left[ \sum_{x_j \neq x_i}\overline{\psi}_{x_i,x_j,0}(\nu) \frac{\overline{\phi'}_{x_j}(\nu)}{\overline{\phi}_{x_j}(\nu)}
   + \sum_{x_j,c=\pm 1}\overline{\psi}_{x_i,x_j,c}(\nu) \frac{\overline{\phi'}_{x_j}(\nu)}{\overline{\phi}_{x_j}(\nu)}\right] P_{(x_j,J)}(t_0),
\end{eqnarray}
hence,
\begin{eqnarray}
\fl \tilde{\overline{I}}_{x_i}(s,\nu)=
    \left[ \sum_{x_j \neq x_i}\overline{\psi'}_{x_i,x_j,0}(\nu) 
   + \sum_{x_j,c=\pm1} \e^{-cs} \overline{\psi'}_{x_i,x_j,c}(\nu) \right] \tilde{P}_{x_j}(s,t_0)\nonumber \\
\fl \quad -\left[ \sum_{x_j \neq x_i}\overline{\psi'}_{x_j,x_i,0}(\nu)
   + \sum_{x_j,c=\pm1}\overline{\psi'}_{x_j,x_i,c}(\nu) \right]  \tilde{P}_{x_i}(s,t_0) \nonumber \\
\fl \quad + \left[ \sum_{x_j \neq x_i}\overline{\psi}_{x_j,x_i,0}(\nu) \frac{\overline{\phi'}_{x_i}(\nu)}{\overline{\phi}_{x_i}(\nu)}
   +  \sum_{x_j,c=\pm1} \e^{-cs} \overline{\psi}_{x_j,x_i,c}(\nu) \frac{\overline{\phi'}_{x_i}(\nu)}{\overline{\phi}_{x_i}(\nu)} \right] \tilde{P}_{x_i}(s,t_0) \nonumber \\
\fl \quad -\left[ \sum_{x_j \neq x_i}\overline{\psi}_{x_i,x_j,0}(\nu) \frac{\overline{\phi'}_{x_j}(\nu)}{\overline{\phi}_{x_j}(\nu)}
   + \sum_{x_j,c\pm1} \psi_{x_i,x_j,c}(\nu) \frac{\overline{\phi'}_{x_j}(\nu)}{\overline{\phi}_{x_j}(\nu)}\right] \tilde{P}_{x_j}(s,t_0).
\end{eqnarray}
As at the beginning of the observation time the total current is zero,
we can replace $\tilde{P}_{x_0}(s,t_0)$ with $P_{x_0}(t_0)$.
Using the WTDs \eref{eq:WTD_alpha_1} or \eref{eq:WTD_alpha_stable}
we again find that the limit as $\nu \to 0$ is finite,
hence also $\tilde{I}_{x_i}(s,t)$ decays to zero in the long-time limit.

The initial-condition term may still substantially affect the
large deviation functionals and their numerical evaluation,
as such a decay may be slow for certain choices of WTDs.
In general, $\tilde{\overline{I}}_{x_i}(s,\nu)$
does \textit{not} vanish even when $\overline{\psi'}_{x_i,x_j,c}(\nu)=\overline{\psi}_{x_i,x_j,c}(\nu)$.
In fact, in this case we have
\begin{eqnarray}
\hspace*{-1.6cm} \tilde{\overline{I}}_{x_i}(s,\nu) =& \sum_{x_i} \left\{   (\e^{-s} - 1)\left[  \overline{\psi}_{x_j,x_i,+1}(\nu) P_{x_i}(s,t_0) + \overline{\psi}_{x_i,x_j,+1}(\nu)P_{x_j}(s,t_0) \right] \right. \nonumber \\
\fl &+ \left. (\e^{s} - 1)\left[ \overline{\psi}_{x_j,x_i,-1}(\nu) P_{x_i}(s,t_0) +\overline{\psi}_{x_i,x_j,-1}(\nu) P_{x_j}(s,t_0) \right] \right\},
\end{eqnarray}
which is in general non-zero (except for $s=0$, when $\overline{I}_{x_i}(\nu)=0$ is recovered).
Consequently, the algorithm of section \ref{sec:The_cloning_step} of the main text must
be iterated for sufficiently long time in order to neglect this finite-time contribution.
We finally mention that, for exponentially distributed waiting times, 
i.e., $\overline{\psi}_{x_i,x_j,c}(\nu)=\beta_{x_i,x_j,c} / (\beta_{x_j} + \nu)$
and $\overline{\psi'}_{x_i,x_j,c}(\nu)=\beta'_{x_i,x_j,c} / (\beta'_{x_j} + \nu)$,
the finite-time effects are minor, as shown by the exact equation
\begin{eqnarray}
\fl \tilde{\overline{I}}_{x_i}(s,\nu)
= \sum_{x_j \neq x_i}  \frac{\tilde{\beta}_{x_j,x_i,0} - \tilde{\beta'}_{x_j,x_i,0}}{\tilde{\beta'}_{x_i} + \nu} P_{x_i}(s,t_0) + 
\sum_{x_j,c=\pm1} \frac{\tilde{\beta}_{x_j,x_i,c} - \tilde{\beta'}_{x_j,x_i,c}}{\tilde{\beta'}_{x_i} + \nu} P_{x_i}(s,t_0)  \nonumber \\
\fl \quad +  \sum_{x_j \neq x_i} \frac{\tilde{\beta'}_{x_i,x_j,0} - \tilde{\beta}_{x_i,x_j,0}}{\tilde{\beta'}_{x_j} + \nu} P_{x_j}(s,t_0) +
\sum_{x_j,c=\pm1} \frac{\tilde{\beta'}_{x_i,x_j,c} - \tilde{\beta}_{x_i,x_j,c}}{\tilde{\beta'}_{x_j} + \nu} P_{x_j}(s,t_0)  ,
\end{eqnarray}
which implies an exponential decay of $\tilde{I}_{x_i}(s,t)$ to zero.

\section{Discrete-time case}
A discrete-time chain can be seen as a stochastic process in continuous time
where the next jump occurs after a constant waiting time of one unit.
Such a scenario can be represented by means of a process  with WTDs  $\psi_{x_n,x_{n-1}}[\tau; w(t)] = p_{x_n,x_{n-1}}[\tau;w(t)] \psi_{x_{n-1}}(\tau)$,
where $p_{x_n,x_{n+1}}[\tau;w(t)]$ is an entry of a transfer matrix and $\psi_{x_{n-1}}(\tau) = \delta(\tau-1)$ is the Dirac delta measure translated by $1$. 
In fact, the procedure of~\sref{sec:The_cloning_step} can be implemented with reasonable accuracy by setting
$\psi_{x_{n-1}}(\tau) = (\sigma \sqrt{2 \pi} )^{-1}\exp \left[ -(\tau-1)^2/(2 \sigma^2)\right]$, with $\sigma \ll 1$.
A discrete-time \textit{Markov} chain can be seen as a special DTI semi-Markov process,
since the transition probabilities do not depend on $w(t)$.
However, such a continuous-time implementation neglects the major computational advantage of dealing with discrete time,
namely, all the ensemble elements can be updated simultaneously. Therefore, we suggest the following parallel algorithm:
\begin{enumerate}[label=\arabic*)]
\item Set up an ensemble of $N$ clones and initialise each to its own random configuration $x_0$.
Also, initialise a unique counter to $n=1$, the variable $C$ to zero, and each element of an
array $\mathbf{C}$ of length $N$ to $1$.
\item For each clone, update the configuration from $x_{n-1}$ to $x_{n}$ according to the mass $p_{x_n,x_{n-1}}[1;w(n-1)]$.
Store the individual values of $\e^{-s \theta_{x_n, x_{n-1}}}$ in $\mathbf{C}$.
\item \textbf{Cloning step.} \label{Sitem} Compute the arithmetic mean  $y$  of all the entries of $\mathbf{C}$.
Perform a weighted random sampling with repetition (see, e.g.,~\cite{SEfraimidis2006}) of $N$ clones from the ensemble, according to their weights $\mathbf{C}$.
This sample replaces the existing ensemble.
\item Increment $C$ to $ C + \ln (y)$. Update $n$ to $n+1$ and reiterate from 2, until $n$ reaches the desired simulation time.
\end{enumerate}
The SCGF is recovered as $-C/n$ for large $n$ (results not shown).
As the sampling at step~\ref{Sitem} is performed simultaneously for all the clones,
it is very unlikely for a single clone to replace all the remaining ones, even in the presence of a strong bias.
This further reduces the finite ensemble effects.

Finally, to make the link to the procedure proposed in~\cite{SGiardina2006}, it is worth
noting that, for the Markovian case, we can arrange the biased WTD as
\begin{equation}
    \tilde{\psi}_{x_i,x_j,c}(\tau) =  \frac{ \sum_{x_k,c'} \tilde{\beta}_{x_k,x_j,c'}}{ \sum_{x_k,c'} \beta_{x_k,x_j,c'}}
    \frac{\tilde{\beta}_{x_i,x_j,c}}{ \sum_{x_k,c'} \tilde{\beta}_{x_k,x_j,c' }  }  \delta(\tau-1),
\end{equation}
where we implicitly assume $\tilde{\beta}_{x_j,x_j,0}=\beta_{x_j,x_j,0} = 0$.
This suggests  the following steps for each ensemble element: increase the time by one unit,
change the state according to the modified transition probability $\tilde{\beta}_{x_i,x_j,c}/\sum_{x_k,c'} \tilde{\beta}_{x_k,x_j,c'}$
and modify the ensemble population according to a cloning factor
$ \sum_{x_k,c'} \tilde{\beta}_{x_k,x_j,c'}/\sum_{x_k,c'} \beta_{x_k,x_j,c' }$, as indeed explained in~\cite{SGiardina2006}.

\section{Model with hidden variables}
In general, non-exponential waiting times arise  when the system configuration at
the present time does not uniquely determine the probabilities at future times.
In this case, we can think that such a configuration is only an incomplete description,
which needs further information about the history and the age (i.e., \textit{memory})
in order to assign the probability of future states.
However, there are situations in which such information can be simply encoded into
additional states, thus extending the configuration space, but permitting a Markovian description.
Such additional states are referred to as \textit{phases} (or \textit{stages}) and said to be \textit{hidden}.
Generically, probability distributions that define waiting times with such a property are
referred to as \textit{phase-type} distributions~\cite{Sstewart2009probability}.
A typical example is the Gamma distribution with integer shape $k$ (also called Erlang distribution)
which describes the random time that a Markovian walker needs to escape $k$ exponential phases in series.
Rather than thinking of a system that leaves its visible configuration after a non-exponential waiting time,
we assume that the system jumps through a set of phases with exponential waiting times, before arriving to the next
visible configuration.

It is proved in~\cite{Scox1955} that any probability density distribution having
a rational Laplace transform $\overline{f}(\nu)$,
with $k$ poles and numerator of degree at most $k$, can be reproduced by a
sequence of $k$ exponential phases. Probability distributions with this property
are called \textit{Coxian} and are representations of certain phase-type distributions.
Without loss of generality, we can make the following partial fraction decomposition
\begin{equation}
  \overline{f}(\nu) = p_0 + q_0 p_1 \frac{\lambda_1}{\nu+\lambda_1} +\sum^k_{i=2} q_0 \ldots q_{i-1} p_i \prod_{l=1}^i \frac{\lambda_l}{\nu+\lambda_l},
  \label{eq:laplace_cox}
\end{equation}
where  the poles are at $-\lambda_i$, $i=1,2,\ldots,k$, $p_{i-1}+q_{i-1}=1$ and $p_k=1$.
Equation~\eref{eq:laplace_cox} has a simple interpretation in the time domain. At each stage $i-1$, there is a
probability $p_{i-1}$ of immediate escape and a probability $q_{i-1}$ of entering the  stage $i$, whose
WTD is exponential with rate $\lambda_i$.

\begin{figure}
\includegraphics[width=0.5\textwidth]{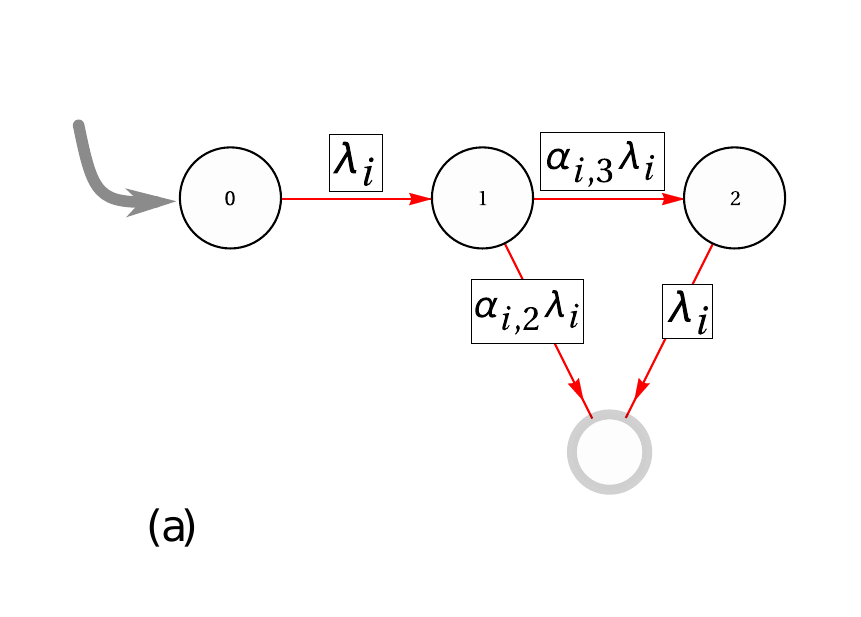}
\includegraphics[width=0.5\textwidth]{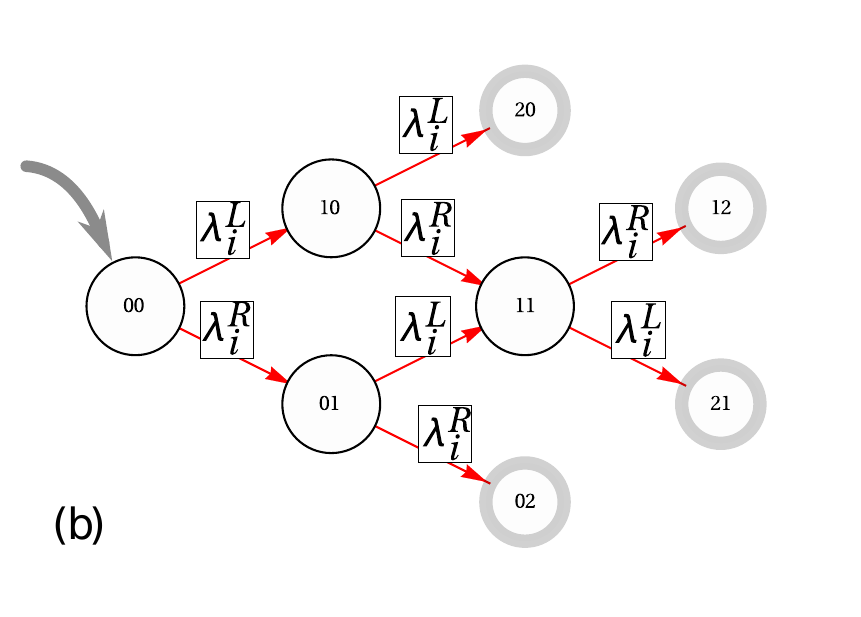}
    \caption{   \label{Sfig:WTD}
Two graphical representations of the WTD \eref{Seq:convenientWTD}.
The waiting time is equal to the adsorption time of a random walker
from the leftmost site to any of the grey sites.  
  }
\end{figure}

We turn now our attention to the model defined by the WTDs~\eref{eq:semiMnoDITWTD1}
and \eref{eq:semiMnoDITWTD2} of the main text.
The Laplace transform of the total WTD $\psi_{x_j,x_i}(t)$, with $i=1,2$, is
\begin{equation}
\overline{\psi}_{x_j,x_i}(\nu)  
     = \frac{(\lambda^L_i)^2 (\nu + 3\lambda^R_{i}+\lambda^L_{i} )}{ \left(\nu + \lambda^R_{i}+\lambda^L_{i}\right)^3}
      + \frac{(\lambda^R_i)^2 (\nu + \lambda^R_{i}+3\lambda^L_{i} ) }{\left(\nu + \lambda^R_{i}+\lambda^L_{i}\right)^3},
    \label{eq:WTD_LAPLACE_noDTI}
\end{equation}
which is a rational function of $\nu$;
its first term corresponds to the right boundary,
while the second one corresponds to the left boundary.
Notice that there is no dependence on the arrival state $x_j$,
the model being defined on a two-state configuration space.
Equation \eref{eq:WTD_LAPLACE_noDTI} can be conveniently written as
\begin{equation}
\overline{\psi}_{x_j,x_i}(\nu )= 
      \alpha_{i,2} \frac{ (\lambda^R_{i}+\lambda^L_{i})^2}{(\nu + \lambda^R_{i}+\lambda^L_{i})^2} 
+      \alpha_{i,3} \frac{ (\lambda^R_{i}+\lambda^L_{i})^3}{(\nu + \lambda^R_{i}+\lambda^L_{i})^3},
\label{Seq:convenientWTD}
\end{equation}
with
\begin{eqnarray}
    \alpha_{i,2} &= \frac{(\lambda^R_{i})^2+(\lambda^L_{i})^2}{\left( \lambda^R_{i}+\lambda^L_{i}\right)^2},\\
    \alpha_{i,3} &= 1 - \frac{(\lambda^R_{i})^2+(\lambda^L_{i})^2}{\left(\lambda^R_{i}+\lambda^L_{i}\right)^2} = \frac{2 \lambda^R_{i} \lambda^L_{i}}{\left(\lambda^R_{i}+\lambda^L_{i}\right)^2},
\end{eqnarray}
thus clearly defining a Coxian distribution.
To separate the effect of boundaries we separately decompose in partial fractions
the left and right WTD contributions of equation~\eref{eq:WTD_LAPLACE_noDTI}, i.e.,
\begin{eqnarray}
    \overline{\psi}_{x_j,x_i}(\nu)
=&   \frac{ \left( \lambda^L_i \right)^2 2 \lambda^R}{\left(\nu+\lambda^R_{i}+\lambda^L_{i}\right)^3} + \frac{ \left( \lambda^L_i \right)^2 }{\left(\nu+\lambda^R_{i}+\lambda^L_{i}\right)^2}  \nonumber \\
&+   \frac{ \left( \lambda^R_i \right)^2 2 \lambda^L}{\left(\nu+\lambda^R_{i}+\lambda^L_{i}\right)^3} + \frac{ \left( \lambda^R_i \right)^2 }{\left(\nu+\lambda^R_{i}+\lambda^L_{i}\right)^2},
\end{eqnarray}
which  can be rearranged as
\begin{eqnarray}
    \overline{\psi}_{x_j,x_i}(\nu) =& \alpha_{i,2}^{L} \frac{\left(\lambda^R_{i}+\lambda^L_{i}\right)^2}{\left(\nu+\lambda^R_{i}+\lambda^L_{i}\right)^2} + \alpha_{i,3}^{L} \frac{\left(\lambda^R_{i}+\lambda^L_{i}\right)^3}{\left(\nu+\lambda^R_{i}+\lambda^L_{i}\right)^3} \nonumber \\
    &  + \alpha_{i,2}^{R} \frac{\left(\lambda^R_{i}+\lambda^L_{i}\right)^2}{\left(\nu+\lambda^R_{i}+\lambda^L_{i}\right)^2} +  \alpha_{i,3}^{R}  \frac{\left(\lambda^R_{i}+\lambda^L_{i}\right)^3}{\left(\nu+\lambda^R_{i}+\lambda^L_{i}\right)^3},
\end{eqnarray}
where
\begin{eqnarray}
\alpha^{L}_{i,2}  = \frac{ \left( \lambda^L_i \right)^2}{ \left( \lambda^R_i+\lambda^L_i \right)^2 } , \quad
\alpha^{L}_{i,3} &= \frac{2\left( \lambda^L_i \right)^2 \lambda^R_i}{\left( \lambda^R_i+\lambda^L_i \right)^3}  , \label{eq:WTD_LAPLACE_noDTI_left_right1} \\
\alpha^{R}_{i,2}  = \frac{ \left( \lambda^R_i \right)^2}{ \left( \lambda^R_i+\lambda^L_i \right)^2}, \quad
\alpha^{R}_{i,3} &= \frac{2\left( \lambda^R_i \right)^2 \lambda^L_i}{\left( \lambda^R_i+\lambda^L_i \right)^3}.
\label{eq:WTD_LAPLACE_noDTI_left_right2}
\end{eqnarray}
Notice that $\alpha_{i,2}^{L} +\alpha_{i,3}^{L}+\alpha_{i,2}^{R}+\alpha_{i,3}^{R} =1$.
The first and second terms correspond to left jumps, while the third and fourth
terms correspond to right jumps.
We also underline that the choice~\eref{eq:WTD_LAPLACE_noDTI_left_right1} and \eref{eq:WTD_LAPLACE_noDTI_left_right2}
is only one of the possible decompositions of the WTD~\eref{eq:WTD_LAPLACE_noDTI}.
Equation~\eref{eq:semiMnoDTIWTD_Laplace} of the main text follows straightforwardly.
A comparison with~\eref{eq:laplace_cox} shows that it corresponds to the case with three stages (i.e., $k=3$),
$p_0=p_1=0$, and $p_2=\alpha^{R}_{i,2} +\alpha^{L}_{i,2}$.
Hence, the jump from $x_i$ to $x_j$ can be modelled as a process of three stages,
in each of which the system is trapped for an exponentially distributed time with rate $\lambda_i$.
At time zero, with probability $1$, the system enters the first stage and waits there.
Then, again with probability $1$, it enters a second identical stage.
After leaving the second stage, the escape occurs  immediately with probability $p_2$,
or the system enters the third and last phase with probability $1-p_2$.
Hence the WTD is the time to absorption of the Markov process with the transition graph of figure~\ref{Sfig:WTD}(a),
given that we start at state $0$.
Recalling the notion of trigger, it is possible to build an alternative but equivalent absorbing Markov process
with the same time to absorption. We think of each of the two Gamma triggers ($R$ or $L$) as a device with two exponential stages
(with rate $\lambda^R_i$ or $\lambda^L_i$). The escape occurs when either of the two triggers leaves the last stage.
The transition graph of the associated Markov processes is shown in figure~\ref{Sfig:WTD}(b).

\begin{figure}
\includegraphics[width=0.5\textwidth]{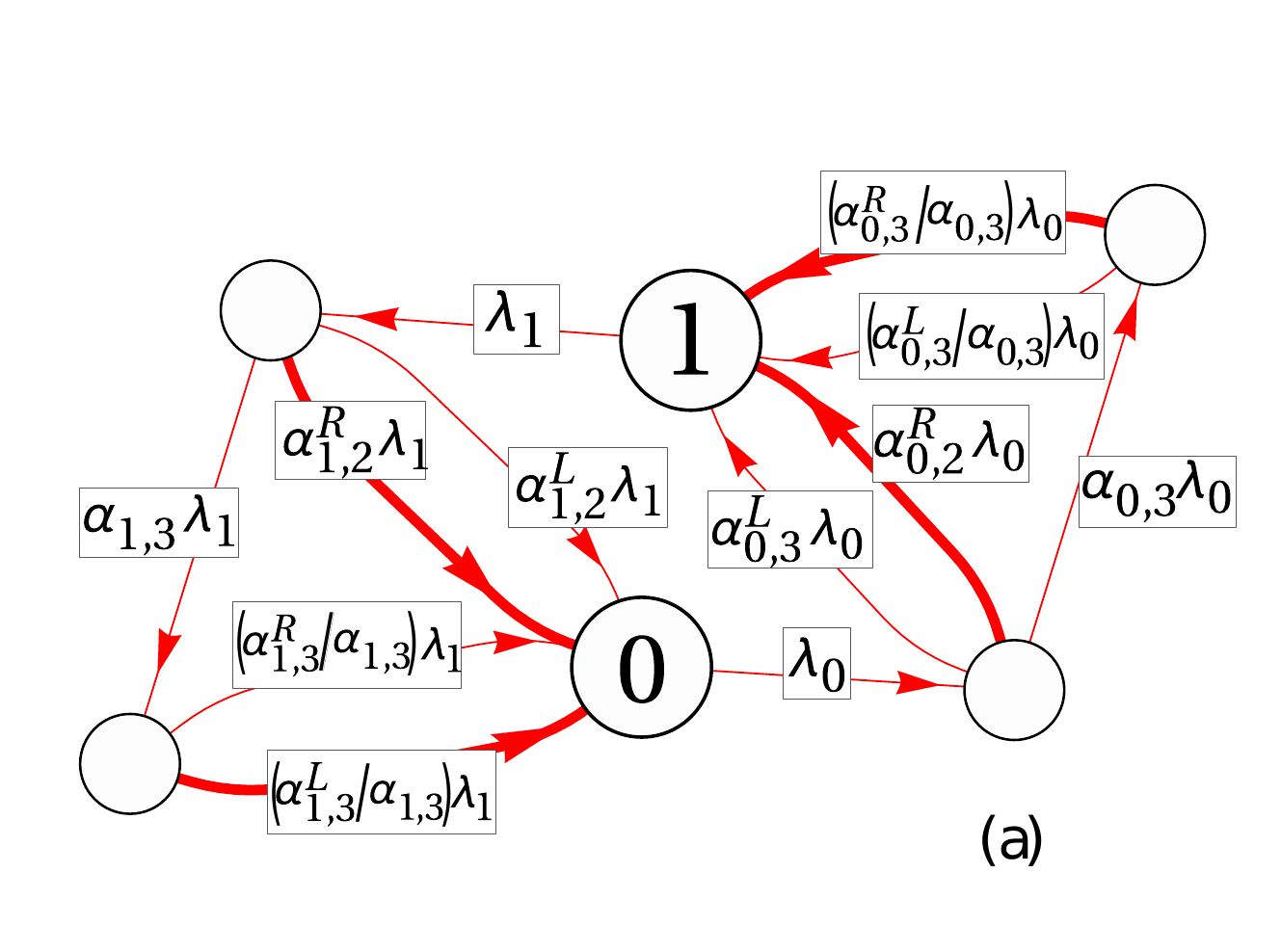}
\includegraphics[width=0.5\textwidth]{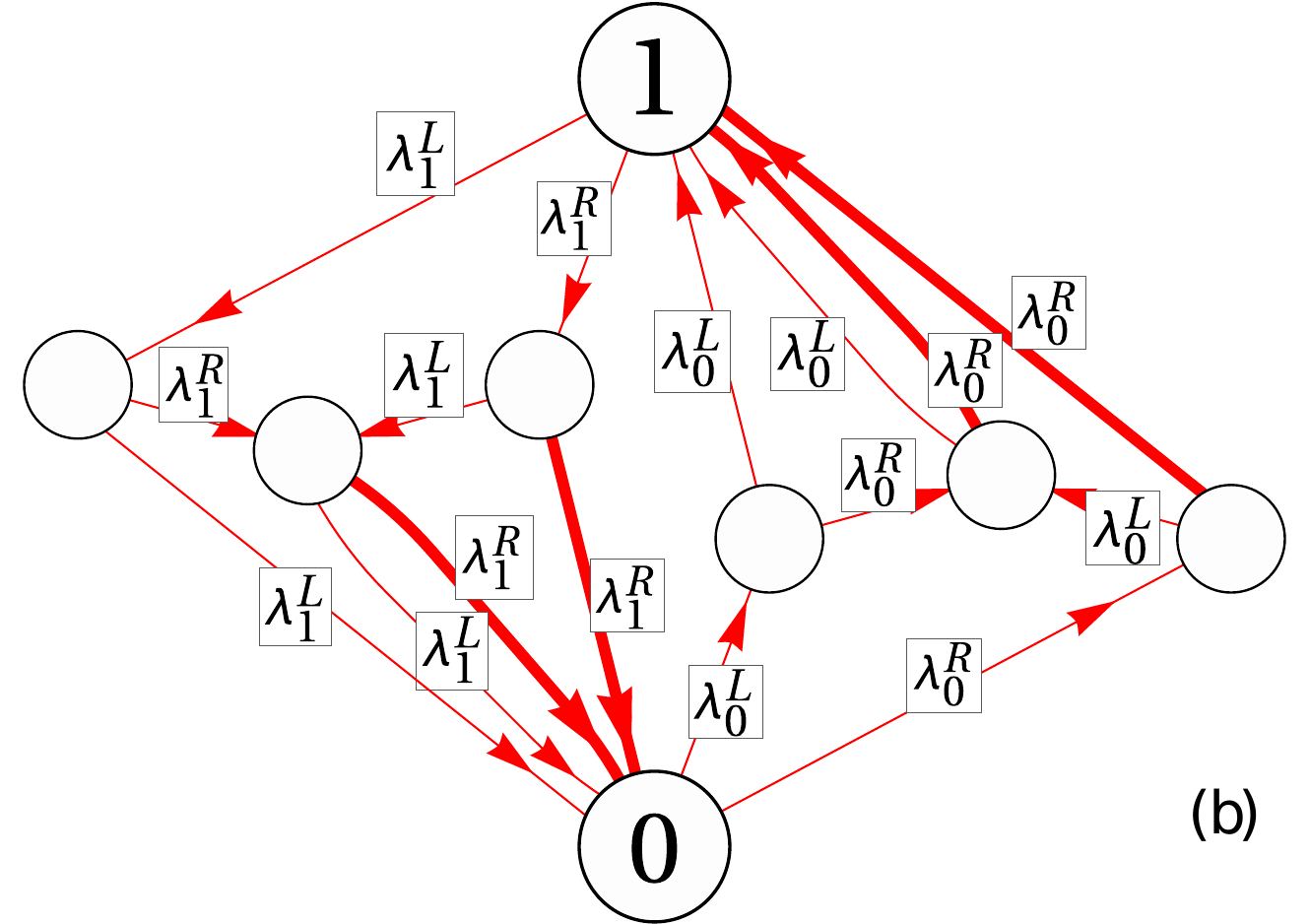}
    \caption{   \label{Sfig:Markovmultigraphs}
Graphical representations of the non-DTI ion-channel model with hidden states.
The bonds corresponding to biased rates are drawn in thick lines.
The modified generators associated with these two models have the
same leading~eigenvalue.
(a) and (b) correspond to the WTD representations of figure~\ref{Sfig:WTD}.
  }
\end{figure}

With the phase-type representations of $\psi_{x_1,x_0}(\tau)$ and $\psi_{x_0,x_1}(\tau)$,
it is straightforward to build a Markov transition graph of the full model.
In order to study the non-equilibrium aspects, we need to distinguish  the contributions of the two boundaries $L$ and $R$.
Hence, in order to obtain the $s$-modified generator and find the SCGF of~\fref{fig:two_state_model}(b) of the main text,
we only bias the true/visible transitions of type $L$.
The resulting Markov representations are then encoded in the multi-graphs of~\fref{Sfig:Markovmultigraphs}.

\end{document}